\documentclass[twocolumn,times,tighten, astrosymb, trackchanges]{aastex631}

\definecolor{bibblue}{cmyk}{1,1,0,0}
\usepackage{amsmath}
\usepackage{float}
\usepackage[caption=false]{subfig}

\graphicspath{{./}{Figures/}}
\begin{document}

\title{Host star properties of hot, warm and cold Jupiters in the solar neighborhood from \textit{Gaia} DR3: clues to formation pathways}
\correspondingauthor{Bihan Banerjee}
\email{banerjeebihan@gmail.com}

\author[0000-0001-8075-3819]{Bihan Banerjee}
\affiliation{Tata Institute of Fundamental Research, Mumbai, 400005, India}
\author[0000-0002-0554-1151]{Mayank Narang}\affiliation{Academia Sinica Institute of Astronomy \& Astrophysics, 11F of Astro-Math Bldg., No.1, Sec. 4, Roosevelt Rd., Taipei 10617, Taiwan, R.O.C.}
\affiliation{Tata Institute of Fundamental Research, Mumbai, 400005, India}
\author[0000-0002-3530-304X]{P. Manoj}
\affiliation{Tata Institute of Fundamental Research, Mumbai, 400005, India}
\author[0000-0002-1493-300X]{Thomas Henning}\affiliation{Max-Planck-Institut für Astronomie (MPIA), Königstuhl 17, D-69117 Heidelberg, Germany}
\author[0000-0002-9497-8856]{Himanshu Tyagi}
\affiliation{Tata Institute of Fundamental Research, Mumbai, 400005, India}
\author[0000-0002-9967-0391]{Arun Surya}
\affiliation{Tata Institute of Fundamental Research, Mumbai, 400005, India}
\affiliation{Indian Institute of Astrophysics, Bangalore}
\author[0000-0002-4638-1035]{Prasanta K. Nayak}
\affiliation{Tata Institute of Fundamental Research, Mumbai, 400005, India}
\affiliation{Instituto de Astrof{\'i}sica, Pontificia Universidad Cat{\'o}lica de Chile, Avenida Vicu{\'n}a Mackenna 4860, 7820436, Macul, Santiago, Chile}
\author[0009-0007-2723-0315]{Mihir Tripathi}
\affiliation{Tata Institute of Fundamental Research, Mumbai, 400005, India}
\affiliation{Academia Sinica Institute of Astronomy \& Astrophysics, 11F of Astro-Math Bldg., No.1, Sec. 4, Roosevelt Rd., Taipei 10617, Taiwan, R.O.C.}

\begin{abstract}
Giant planets exhibit diverse orbital properties, hinting at their distinct formation and dynamic histories. In this paper, using \textit{Gaia} DR3, we investigate if and how the orbital properties of Jupiters are linked to their host star properties, particularly their metallicity and age. We obtain metallicities for main sequence stars of spectral type F, G, and K, hosting hot, warm, and cold Jupiters with varying eccentricities. We compute the velocity dispersion of host stars of these three groups using kinematic information from \textit{Gaia} DR3 and obtain average ages using velocity dispersion-age relation. We find that host stars of hot Jupiters are relatively metal-rich ([Fe/H]=$0.18 \pm 0.13$) and young ( median age $3.97 \pm 0.51$ Gyr) compared to the host stars of cold Jupiters in nearly circular orbits, which are relatively metal-poor ($0.03 \pm 0.18$) and older (median age $6.07 \pm 0.79$ Gyr). Host stars of cold Jupiters in high eccentric orbits, on the other hand, show metallicities similar to that of the hosts of hot Jupiters, but are older, on average (median age $6.25 \pm 0.92$ Gyr).
The similarity in metallicity between hosts of hot Jupiters and hosts of cold Jupiters in high eccentric orbits supports high eccentricity migration as the potential origin of hot Jupiters, with the latter serving as the progenitors. However, the average age difference between them suggests that the older hot Jupiters may have been engulfed by the star in a timescale of $\sim 6$ Gyr. This allows us to estimate the value of stellar tidal quality factor $Q'_\ast\sim10^{6\pm1}$. 

\end{abstract}

\section{Introduction} \label{sec:intro}
Outside our solar system, more than 5000 planets have been detected and the number increases every day. The properties of the discovered exoplanets are often very different from those of the solar system planets. For example, many of the first detected exoplanets were hot Jupiters, i.e., giant planets with an orbital period shorter than 10 days \citep[e.g.,][]{Mayor1995, Butler1997}. 

Planet properties, though seemingly diverse and different, are found to be tightly correlated with their host star properties \citep[e.g.,][]{Hsu2019, Yang2020, CKS4, Narang18, Mulders2018, Gaudi2021, ZhuandDong2021}. For example, stellar spectral type and stellar mass, and accompanying planet mass and size are found to be correlated \citep[e.g.,][]{Howard2010, Howard2012, Fressin2013, Dressing2015, Mulders2015, Hardegree-Ullman2019, Yang2020}. Another robust correlation exists between host star metallicity and planet mass \citep[e.g.,][]{Gonzalez1997, Santos2003, Fischer2005, Santos2006, Urdy_Santos2007, Reffert2015, Santos17, Narang18}. The occurrence rate of giant planets increases with host star metallicity, and on average Jupiter-hosting stars are likely to be of super-solar metallicity ([Fe/H]$\sim 0.18\pm0.05$) \citep[e.g.,][]{Narang18, Mulders2018, CKS4}. However, many of these results are derived from the Kepler sample which is complete for planets with orbital periods $\leq 1$ year \citep[e.g.,][]{Narang18, CKS4}. Occurrence rate studies with Kepler data reveal that although the short-period Jupiters are easier to detect, they are much rarer than the small planets with similar orbital periods \citep[e.g.,][]{CKS4, Hsu2019, Mulders2018, Gaudi2021}. Recently many more warm Jupiters have been detected with Transiting Exoplanet Survey Satellite (\textit{TESS}) \citep[e.g.,][]{Eberhardt2023, Lubin2023}, and long-period Jupiters have been discovered with radial velocity of higher precision.  Occurrence rate studies with combined data of transit and radial velocity surveys show that giant planet occurrence rate increases beyond 1 au and peaks around an orbital distance of $\sim3$ au \citep[e.g.,][]{Fernandes18, Fulton21, Kumimoto21, Wolthoff2022}. The most well-studied gas-giant, Jupiter in the solar system, is located at a distance of $5.2$ au from the sun and orbits a star with relatively lower metallicity ([Fe/H]$=0$) than the hot Jupiter hosts ([Fe/H]$\approx 0.18$ dex) \citep{Narang18, Mulders2018, CKS4}. Very long-period Jupiters, observed through the direct-imaging method, are found to show no particular preference towards host-star metallicity, and the average metallicity is around solar to sub-solar \citep[e.g.,][]{Swastik21}. Several high-resolution radial velocity surveys report that the average metallicity of long-period Jupiter hosts is close to the solar value \citep[e.g.,][]{Fulton21, Wolthoff2022}. However, not all long-period Jupiters are Jupiter analogs. A large fraction of the long-period Jupiters detected through RV observations have large orbital eccentricity \citep{Bitsch2020, NadiaFord2011, CKSIV2024}, unlike the solar system Jupiter. On the other hand, orbits of short-period Jupiters are predominantly circular \citep[e.g.,][]{Jackson2023}. Do these orbital diversities of Jupiters hint at different formation and evolution scenarios? A strong connection between the orbital architecture of planets and host star properties could provide clues to the formation history. Therefore, it is only natural to ask, \textbf{how do the host star properties correlate with the orbital properties (orbital period and eccentricity) of the giant planets?}

 \cite{Buchhave18} and \cite{Maldonado18} have attempted to answer this question; however, they arrive at slightly different conclusions. \cite{Maldonado18}, with metallicity measured from the high-resolution spectra for a sample of 88  host stars of giant planets, found that hot Jupiter hosts are relatively metal-rich than hosts of cold Jupiters. \cite{Maldonado18} also showed that host stars of cold Jupiters are relatively richer in the $\alpha$ elements than hot Jupiters. They argued that their results suggest a different formation mechanism for hot and cold Jupiters.
 %Analyzing other properties of host stars of hot and cold Jupiters \cite{Maldonado18} suggested cold and hot Jupiters form in different ways. 
 On the other hand, \cite{Buchhave18} performed their analysis with 65 Jupiter hosts, and showed that hosts of both of the hot Jupiters, and the cold Jupiters with high orbital eccentricity, have super-solar metallicity and likely come from the same population; host stars of low-eccentric cold Jupiters have solar or sub-solar metallicity, and belong to a different population.     

The simple question of host star metallicity-orbital period correlation has deep implications for the formation pathways of the Jupiters. It is still unclear how Jupiters are formed at different orbital locations \citep[e.g.,][]{Dawson18}. The increase of giant planet occurrence rate with host star metallicity is best explained by core-accretion theory \citep[e.g.,][]{Bodenheimer1986, Poll96, Ikoma2001, Mordasini2008, Dangelo_Lissauer2018}. According to this theory, a solid core of a critical mass forms first ($\sim 10M_\oplus$ at 5 au in minimum mass solar nebula condition, see also \citep[e.g.,][]{Piso2014, Piso2015}),  which then accretes gas from the disk to form a Jupiter-like planet. However, in all protoplanetary disks, available solid mass at a given radius ($r$) increases with $r$ \citep[e.g.,][]{Armitage2020, Powell2019}. The sticking efficiency of grains also increases beyond the snow line \citep[][]{Ozukumi2012, Armitage2020}. Simulations show the efficiency of forming such massive cores is highest beyond the snowline of the disk, i.e. at $2-5$ au from a sun-like star \citep[e.g.,][]{Mordasini12}. To complete this process within the disk dispersal timescale, a metal-rich disk is necessary \citep[e.g.,][]{IdaLin2005, Kornet2005, WyattClarke2007, Boss2010, Mordasini12}. Since the star and the circumstellar disk form from the same interstellar cloud material, we do expect more giant planets around metal-rich systems. However, even in metal-rich systems, the formation mechanisms of hot and warm Jupiters are still not clearly understood. Two major theories, widely discussed in the literature for the formation of close-in giant planets are:

(1) \textbf{In-situ formation:} A good fraction of the detected giant planets have an orbital period $\leq 1$ year. These planets are called hot and warm Jupiters (The rest of them, with orbital period $>1$ year are called cold Jupiters). In the in-situ formation theory, hot and warm Jupiters form at their present locations \citep[e.g.,][]{Lee14, LeeChiang2017}. Solid grains coagulate to form a massive core in the inner disk. If the core mass crosses a certain threshold value (Typically $\sim10 M_\oplus$), it can accrete gas very rapidly \citep[e.g.,][]{Piso2014, Piso2015}. The main challenge for this mechanism is the lack of available solid mass at the inner disk, with the standard disk conditions. However, an enhanced radial drift of pebbles can provide an additional large fraction of mass \citep{Johansen2017}. Since stellar metallicity traces the total available solid mass of the disk, this mechanism can only be active in extremely metal-rich systems \citep{Dawson18}. \cite{Maldonado18}, found hot Jupiter hosts to be significantly metal-rich than colder Jupiter hosts and argued in favor of the in-situ formation of hot Jupiters. If the in-situ formation indeed is the dominant channel of formation, then we expect to find a gradual decrease in the metallicity of host stars with the orbital period of Jupiter they host, irrespective of their orbital eccentricity. 

(2) \textbf{Migration}: Jupiters are formed beyond the snowline, and then they eventually migrate inward. Two possible migration channels are: 

(2.1) \textbf{\textit{Disk migration}}: In this scenario, the Jupiters forming in the outer disk interact with the gas disk and migrate inwards, exchanging angular momentum and dissipating energy \citep[e.g.,][]{Goldreich1980, Lin1986, D'Angelo2003, Baruteau2014}. The final location of the planet is likely decided by mass loss, tidal interactions, inner disk edge location, and many other factors \citep[][]{Chang2010}{}{}. However, in this migration scenario, there is no clear host star metallicity dependence \citep[e.g.,][]{Armitage2020}, and planets undergoing disk migration do not get excited to highly eccentric orbits \citep{Duffell2015}. If disk migration plays a dominant role in forming close-in Jupiters, we expect the Jupiters to be in circular to low-eccentric orbits, with uniform distribution of host star metallicity regardless of the Jupiter's orbital period \citep[e.g.,][]{Goldreich2003, Duffel2015}.

(2.2) \textbf{\textit{High-eccentric tidal migration (HEM)}}: In this scenario, a fraction of cold Jupiters is excited to some highly eccentric orbits, either by planet-planet scattering \citep[e.g.,][]{Chatterjee2008}, or by Kozai-Lidov cycle between planets \citep[e.g.,][]{Kozai1962, Lidov1962}, or by secular interactions \citep[e.g.,][]{Petrovich2015, Hamers2017}, or by interacting with external bodies (e.g. stellar fly-bys) \citep[e.g.,][]{Shara2016, Xiang2016}, and then tidally circularize to a smaller orbit to become a hot Jupiter. The first three mechanisms are preferable in metal-rich systems where multiple giant planets are expected to form \citep[e.g.,][]{Bitsch2015, Buchhave18}. \citet{Buchhave18} found eccentric cold Jupiter hosts to have similar metallicities as hot Jupiter hosts, which supported this mechanism of formation.  

\begin{figure}[h]
    \centering
    \includegraphics[width=0.45\textwidth]{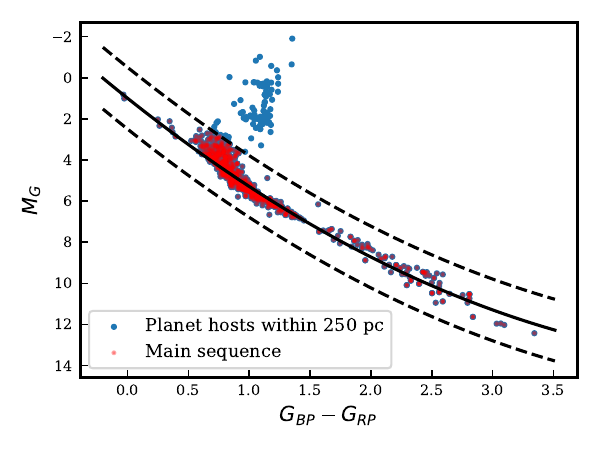}
    \caption{\textit{Gaia} DR3 color-magnitude diagram of planet-host stars in the solar neighborhood. All stars within 250 pc with detected planets are shown as Blue points. Red points indicate the main sequence stars. Black solid line represents Equation \ref{eq:1}, and dotted lines represent $\pm1.5$ magnitude of the expected $M_G$ value for a given $G_{BP}-G_{RP}$ }
    \label{fig:sample1}
\end{figure}

Host star metallicity can help distinguish between these formation mechanisms and hence investigating the host star metallicity-orbital period correlation is vital for improving our understanding of giant planet formation. Although \citet{Buchhave18}  and \citet{Maldonado18} have investigated this correlation, the sample size was smaller, and host star metallicity was derived from different surveys and techniques and therefore was not homogeneous. In addition, as discussed earlier, they arrive at somewhat contradictory conclusions. \citet{Maldonado18} arguing in favour of the in-situ formation and \citet{Buchhave18} finding support for high-eccentric tidal migration. We revisit this problem with a significantly larger sample and homogeneous set of host star metallicities derived from \textit{Gaia} DR3. Metallicities derived from different observations and different instruments tend to suffer from offsets. Therefore, while comparing, it is important to have a homogeneously measured sample of metallicities. In this paper, we have used metallicities of planet hosts uniformly determined from \textit{Gaia} Radial Velocity Spectrograph (RVS). The wavelength range covered by the RVS is 846-870 nm, with medium resolving power R$=\lambda/\Delta\lambda\sim 11500$ \citep{Cropper2018}. Following \textit{GSP-Spec} module \citep{Recio-Blanco16}, chemo-physical parameters of nearly 5.6 million stars of our galaxy have been derived from these spectra and reported in \textit{Gaia} Data release 3 %(hereafter \textit{Gaia}DR3) 
\citep[hereafter \textit{Gaia} DR3;][]{GaiaDR3Part1,Recio-Blanco22}. This sample of stars with homogeneously determined metallicities includes many planet hosts and allows us to compare the host star metallicity of different planet samples. 

Stellar metallicity and stellar age are found to be correlated \citep[e.g.,][]{Carlberg1985, Meusinger1991,nordstrom2004}. As ISM eventually gets enriched in metals, younger stars tend to be born metal-rich. Several works have argued that Jupiter hosts, in addition to being metal-rich are also younger than the field stars and smaller planet hosts \citep[e.g.,][\color{bibblue}{Narang et al. (2024), Under review}]{Swastik21, Mustill2022, Athira2022}. There have been suggestions that Jupiters can only be formed only after a threshold chemical enrichment of ISM has taken place \citep[e.g.,][\color{bibblue}{Narang et al. (2024), Under review}]{Mordasini12, Narang22}. It has been also argued in the literature that hot Jupiter hosts seem younger because older hot Jupiters are getting destroyed by stellar tides \citep[e.g.,][]{Hamer2019, Miyazaki2023}. Therefore, it is an interesting exercise to compare the average metallicity and age of host stars of Jupiters in various orbital distances and eccentricities. \textit{Gaia} DR3 also provides parallax ($\pi$), proper motion in RA and Dec (pmRA, pmDE), and radial velocity information of the stars. Using these measured quantities, velocity dispersion for a group of stars can be obtained which is a proxy for age \citep[e.g.,][]{Binney2000, AumerBinney2009} (See Section \ref{sec: age}). 

In Section \ref{sec:samp} we describe our sample and the selection criteria. Calibration of Gaia metallicities and comparison with metallicities derived from other surveys (e.g. \textit{GALAH, LAMOST}) are discussed in section \ref{sec:cal}. We present our results and discuss their statistical significance in Sections \ref{sec: corr} and \ref{sec: stat}. We discuss the plausible physical origins of the observed results in \ref{sec:dis}, and finally, we summarize in section \ref{sec: sum}. 

\section{Sample selection} \label{sec:samp}
Our analysis requires a sample of main-sequence stars that host giant planets at various orbital distances, with (1) Stellar metallicity ([Fe/H]) homogeneously determined and (2) Radial velocity, parallax, and proper motion measured with acceptable accuracy. \textit{Gaia} DR3 provides a homogeneous dataset for both. 
From the planetary systems table in NASA exoplanet archive \citep{exoarch, Akeson13}, we selected confirmed planets (\texttt{controversial flag=0}) with measured orbital period. This gives us {$4871$} detected planets around {$3612$} host stars. We cross-matched the positions of these planet hosts with \textit{Gaia} DR3 sources (\href{https://vizier.cds.unistra.fr/viz-bin/VizieR-3?-source=I/355/gaiadr3}{I/355/gaiadr3})  and astrophysical parameters catalogs (\href{https://vizier.cds.unistra.fr/viz-bin/VizieR-3?-source=I/355/paramp}{I/355/paramp}) simultaneously, using a search radius of 1 arcsecond. Using the distance measured by \textit{Gaia} parallax \citep{GaiaDR3Part1}, with parallax/parallax error $>10$, we only keep sources within $250$ pc, with robust measurement of distance. 
We want to keep only the main sequence host stars in our sample. \citet{Pecaut13} has provided an  online Table \href{https://www.pas.rochester.edu/~emamajek/EEM_dwarf_UBVIJHK_colors_Teff.txt}{[1]} for standard main-sequence stars which is updated with new photometric observations (Last update was on 16-04-22). In this table, de-reddened colors, and absolute G band magnitude of the standard stars of different spectral types from \textit{Gaia} DR2 are given. Following \citet{Narang22}, we performed a second order polynomial fit between \textit{Gaia} DR2 BP/RP colors and \textit{Gaia} DR2 G band magnitude, restricting ourselves between F4 to M4 spectral types. The best fit solution is shown in Equation \ref{eq:1}(Also see \citep{Narang22}). It has been verified by \citet{Narang22} that the median difference between the values of $\boldsymbol{M_G}$ ,$\boldsymbol{G_{BP}}$ and $\boldsymbol{G_{RP}}$ derived from \textit{Gaia} DR2 and \textit{Gaia} DR3 are 0.01, 0.02 and 0.01 respectively. Therefore, the following Equation holds for \textit{Gaia} DR3 within $\leq 3$ \% . 
\begin{equation}\label{eq:1}
    M_G=-0.43\times(G_{BP}-G_{RP})^2+4.72\times(G_{BP}-G_{RP})+1.00
\end{equation}
\textit{Gaia} DR3, provides apparent magnitude ($m_G$), observed color index ($(G_{BP}-G_{RP})_{obs}$), distances ($d$), the G-band extinction coefficient ($A_G$), and color excess values ($E(BP-RP)$) for all these sources. We note that 99\% of the planet host stars in our sample within 250 pc has an ($A_G$)  value $< 0.5$ , and E(BP-RP) value $< 0.3$. %However, starlight gets dimmed and reddened because of the distance and absorption and scattering by the dust in the Inter Stellar Medium. We need to convert them to absolute magnitude ($\boldsymbol{M_G}$) and intrinsic color $\boldsymbol{(G_{BP}-G_{RP})}$, to use the equation above. The G-band extinction coefficient ($\boldsymbol{A_G}$), and color excess values ($\boldsymbol{E(BP-RP)}$) for each source is also reported in \textit{Gaia} DR3. 
To find the absolute magnitude and intrinsic color,  we use the following equations:
\begin{align}
    M_G &= m_G - 5 \log_{10}{(d-1)}- A_G \\
    (G_{BP}-G_{RP}) &= (G_{BP}-G_{RP})_{obs}-E(BP-RP)
\end{align}
We select stars within $\pm1.5$ magnitude of the expected value of $M_G$ for a given $(G_{BP}-G_{RP})$, following the Equation \ref{eq:1}. %This method of selecting main sequence stars was proposed by \citet{Pecaut13} using \textit{Gaia} DR2 data.% 
In Figure \ref{fig:sample1} all the planet-hosting stars within 250 pc are shown as blue points. The black solid line represents Equation \ref{eq:1}, and two dashed lines are $\pm1.5$ magnitude of the expected $M_G$ value for a given $G_{BP}-G_{RP}$. Therefore the stars within two dashed lines are the main sequence stars and are plotted as red points.    
We find {$610$} main sequence stars hosting {$816$} planets from this sample.

\begin{figure}[htpb]
%\centering
 %\begin{subfigure}{0.45\textwidth}

\subfloat{%
\centering
\includegraphics[width=0.45\textwidth]{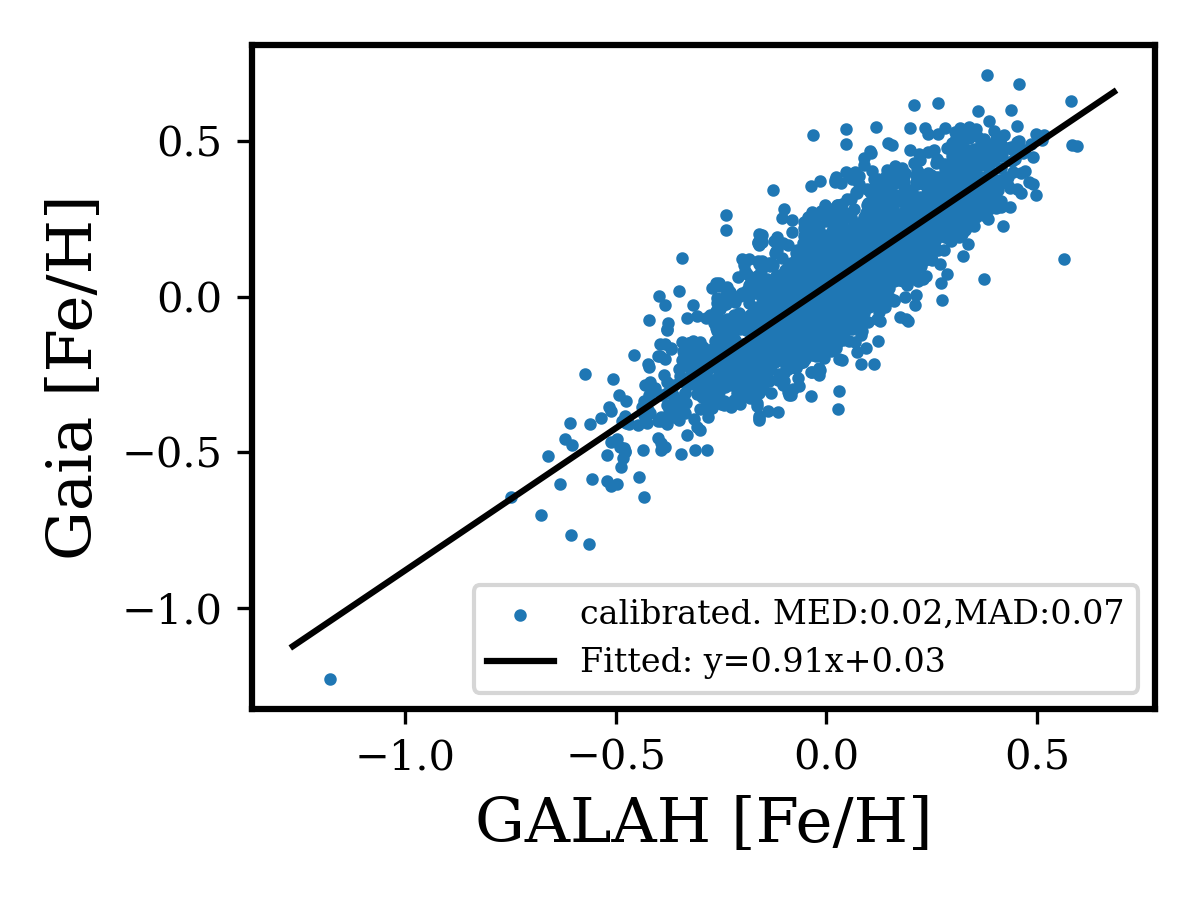}
\label{fig:2a}%
}   
 %\end{subfigure}
 \vfill
 \subfloat{%
     \centering
     \includegraphics[width=0.45\textwidth]{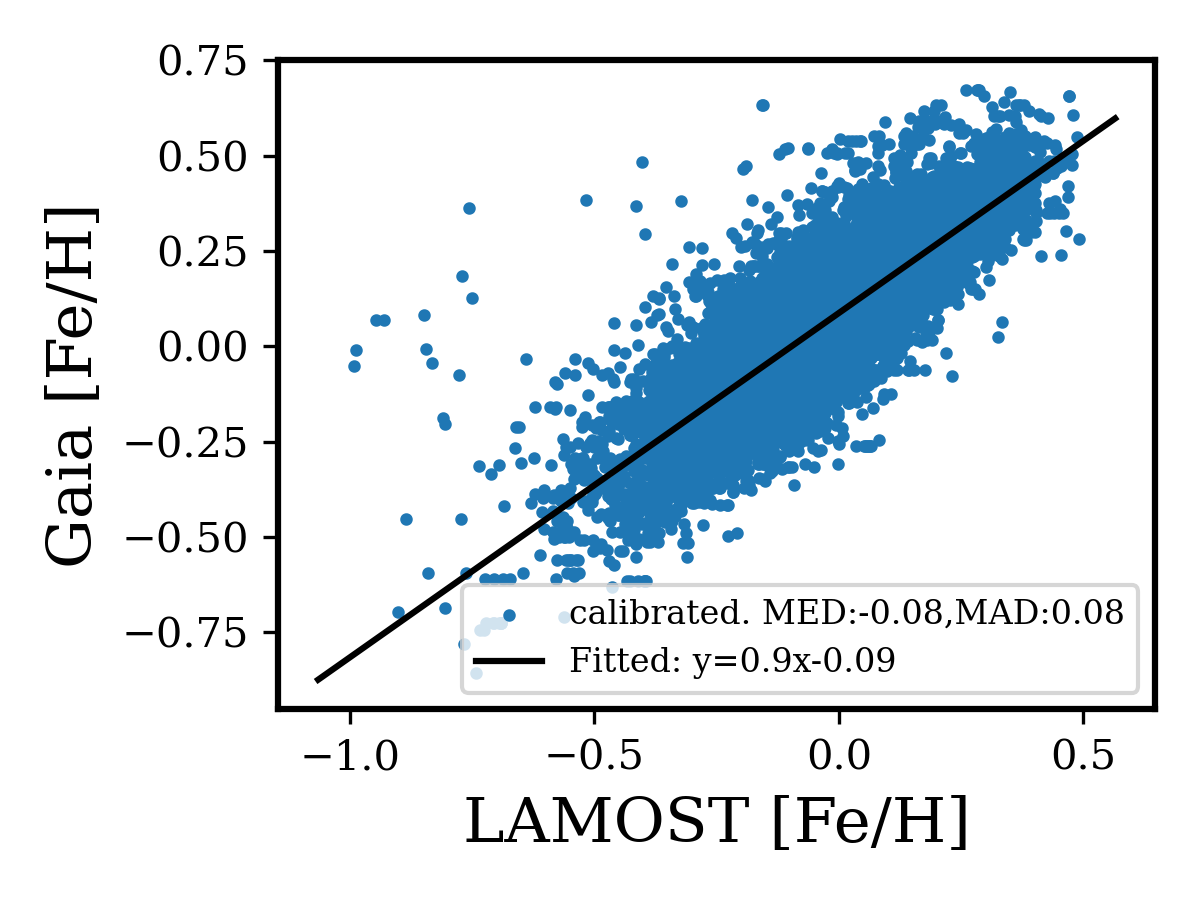}
     \label{fig:2b}%
}     
 %\begin{subfigure}{0.45\textwidth}
\caption{Comparison of the Calibrated \textit{Gaia} [Fe/H] with (a) \textit{GALAH} and (b) \textit{LAMOST} metallicities. The blue points represent the [Fe/H] values of stars common to two instruments. The black solid line is a fitted straight line. MED denotes the median of the difference of [Fe/H] values between (a) \textit{Gaia} \& \textit{GALAH}, and (b) \textit{Gaia} \& \textit{LAMOST}, i.e. the median of (y-x) coordinates of the blue points. MAD denotes the absolute deviation of [Fe/H] differences from this median.}
\label{fig:cal_plot}
\end{figure}
\textit{Gaia} has observed all of these main-sequence, planet-host stars, but the metallicity derived from RVS data is available for only a subset of them. In addition, the quality of RVS data is not the same for all of them within this subset. To select only the best quality data products in [Fe/H], we follow \citet{Recio-Blanco22} and set the first 13 values of the quality flag chain to zero (See Appendix C of \citet{Recio-Blanco22}).  We find $380$ main sequence stars among the $610$ to have [Fe/H] values measured with the highest quality. These $380$ stars host $519$ planets in total. 
%$401$ of them were discovered by radial velocity method, and the remaining $117$ planets are detected in transit.

In this work, we are primarily interested in Jupiters, and their host stars. We define Jupiters as planets with mass ($M\sin i$ or $M$) in between $100M_\oplus$ to $1200 M_\oplus$. From our sample, we restrict ourselves only to planets with mass measured with $3\sigma$ accuracy ($M/\sigma_M\geq 3$), and orbital period measured with $7\sigma$ accuracy ($P/\sigma_P\geq 7$) we end up having $239$ Jupiters, around $209$ host stars. There are 26 systems with multiple Jupiters. 

\begin{figure*}[htbp!]
    \centering
    \includegraphics[width=\linewidth]{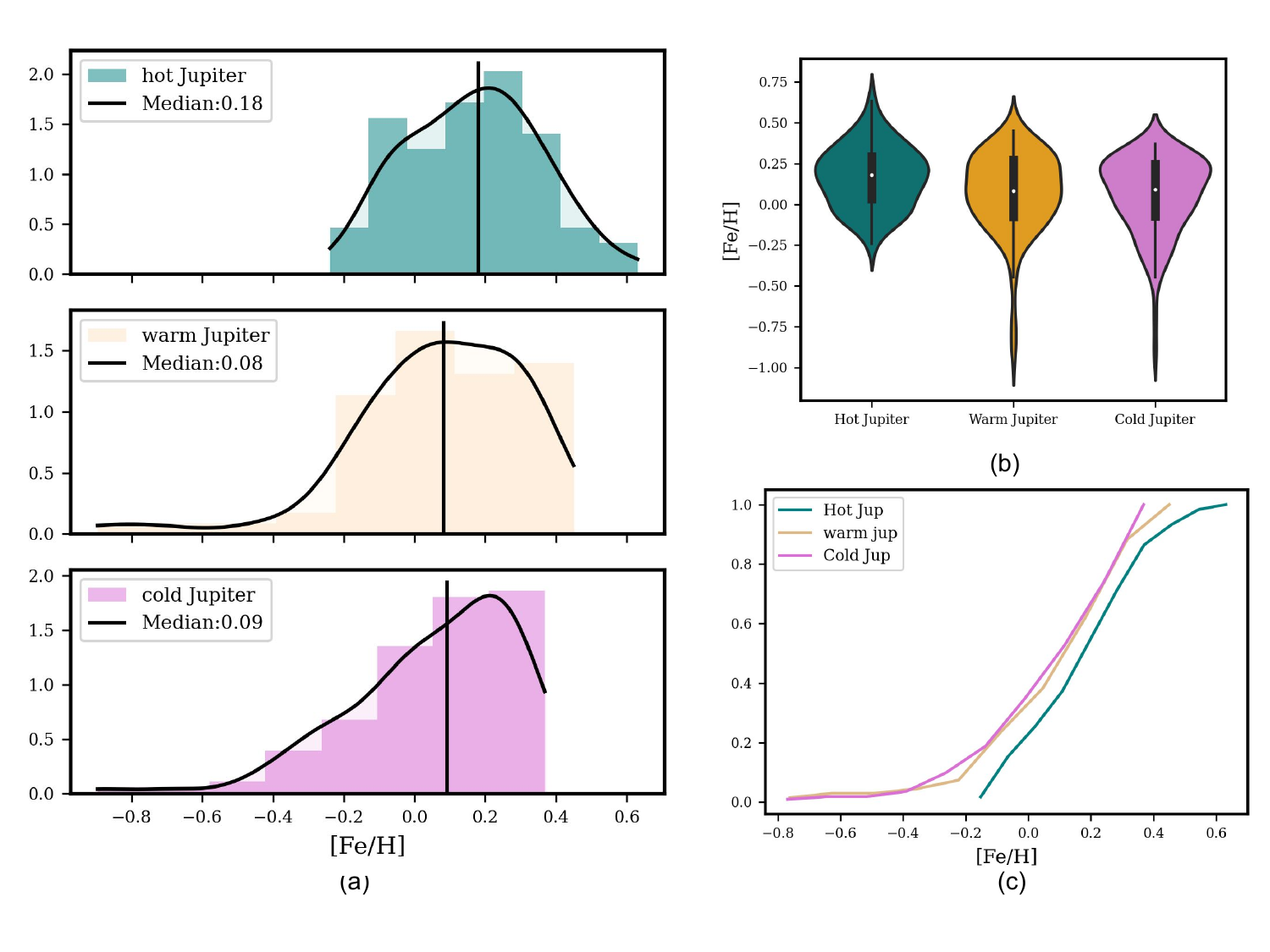}
    \caption{[Fe/H] distributions for hot, warm, and cold Jupiters. (a) Colored histograms and Kernel Density Estimates. Shows HJs are relatively metal-rich compared to warm and cold Jupiter. Any difference between warm and cold Jupiters in terms of host star metallicity is not apparent. The black vertical lines denote the median [Fe/H] (b) Violin plots showing the distribution of [Fe/H] for HJ, WJ, and CJ hosts (c) Cumulative distributions of [Fe/H] for HJ, WJ, and CJ hosts. The plots clearly show hot Jupiter hosts are distinctly metal-rich.}
    \label{fig:3}
\end{figure*}

\section{Calibration of \textit{Gaia} DR3 Metallicities}\label{sec:cal}

\textit{Gaia} DR3 has provided the values of $T_{\text{eff}},\log{g}$, [M/H], [Fe/H] derived from RVS spectra using GSP-Spec MatisseGauguin \citep{Recio-Blanco22}. However, this method consistently finds lower surface gravity ($\log{g}$) values for all the stars, and there is an overall offset compared to the literature. Therefore, a correction needs to be added to recover the true values of $\log{g}$. In addition, there is a weak dependence of metallicity values over $\log{g}$. Hence, a global correction to the [M/H], [Fe/H], [$\alpha$/H] values, by adding a polynomial of uncalibrated $\log{g}$ is suggested by \citet{Recio-Blanco22}. Following the same, we perform the following correction: 
%[M/H]_{calibrated} &=[M/H]_{uncalibrated}+\sum p_n ({\log{g}})^n \\
\begin{align*}
     [Fe/H]_{calibrated} &= [Fe/H]_{uncalibrated} + \sum p'_n({\log{g}})^n
\end{align*}
The values for $p'_n$ are taken from \citet{Recio-Blanco22}. 
%The polynomial coefficients $p_n, p'_n$ for the above two calibrations are given in Table (\ref{tab:tab1}). 
%\begin{center}
 %   \begin{table}[h!]
 %   \centering
 %       \begin{tabular}{ |c|c|c|c|c| } 
 %           \hline
 %Quantity & $p_0$ & $p_1$ & $p_2$ & $p_3$ \\ 
 %\hline
 %$[M/H]$ & 0.274 & -0.1373 & -0.0050 & 0.0048 \\ 
 %$[Fe/H]$ & 0.3699 & -0.0680 & 0.0028 & -0.0004 \\ 
 %           \hline
 %       \end{tabular}
 %       \caption{Polynomial coefficients for calibration \citep{Recio-Blanco22}}
 %       \label{tab:tab1}
 %   \end{table}
%\end{center}
%[Fe/H] values, derived from datasets of separate instruments, tend to suffer from offsets. However, there should be a linear dependence between the two, since both are measuring the same quantity.

After making these corrections to [Fe/H] values from \textit{Gaia} DR3 we compare the calibrated metallicities to corresponding values of \textit{GALAH} DR3 \citep[e.g.,][]{Buder2021} and \textit{LAMOST} DR7 \citep[e,g.,][]{Wang2020}. For this comparison, we select all the main-sequence stars within 250 PC from \textit{Gaia} DR3 (See Section \ref{sec:samp}), with best quality RVS data (first 13 entries of quality flag chain is zero, see \citep[]{Recio-Blanco22}) and cross-match them with \textit{GALAH} DR3 and \textit{LAMOST} DR7 sources within 1 arcsec search radius around each of the \textit{Gaia} DR3 sources. To get the best quality data products of \textit{GALAH} DR3, we follow the standard practice and use the recommended flag settings from GALAH\_DR3\_main\_allstar\_v2.fits catalog: (1) \texttt{snr\_c3\_iraf $> 30$}; (2) \texttt{flag\_sp = 0}, \& \texttt{flag\_fe\_h = 0 }( See \href{https://www.galah-survey.org/dr3/using_the_data/#recommended-flag-values}{[2]}). Between \textit{Gaia} and \textit{GALAH} best quality data, we find 2593 main sequence sources within 250 pc. Similarly, for the best quality data products from LAMOST, we set the [Fe/H] error $<$ 0.3 dex, and two quality flags for R and B band spectral data, bad\_b,bad\_r to zero ( See \href{https://dr7.lamost.org/v2.0/doc/mr-data-production-description}{[3]}). We find 10505 main sequence stars common to \textit{Gaia} DR3 and \textit{LAMOST} DR7 within 250 PC.

Figure \ref{fig:cal_plot} compares calibrated [Fe/H] values from \textit{Gaia} DR3 to that from \textit{GALAH} and \textit{LAMOST} for main sequence stars within 250 pc. The data can be fitted with a straight line with a slope close to 1, indicating that the calibrated stellar metallicity values from \textit{Gaia} DR3 are consistent with \textit{GALAH} and \textit{LAMOST} values of the same. The median [Fe/H] difference between two instruments, i.e. the offset, is small for both cases, 0.02 between \textit{Gaia} DR3 and \textit{GALAH} DR3 and -0.08 between \textit{Gaia} DR3 and \textit{LAMOST} DR7. 
%Ideally, an $y=x$ straight line could be fitted, since both instruments measure the same physical quantity. However, because of instrumental offsets, we expect deviations from this. Nevertheless, the fitted straight lines in both cases have a slope close to 1.

%\section{Analysis}\label{sec:ana}

%\section{Results} \label{sec:res}

\section{Host star metallicity - orbital period connection for Jupiters }\label{sec: corr}

As mentioned in Section \ref{sec:samp}, we have 239 Jupiters around 209 main-sequence stars in our final sample. Metallicities ([Fe/H]) of all these stars are calibrated following the procedure described in Section \ref{sec:cal}. These 239 Jupiters are located at various orbital distances from their host stars.  Based on their orbital period ($P$), these Jupiters can be subdivided into three categories: hot, warm, and cold. We define: (a) \textbf{Hot Jupiters (HJ)} as Jupiters with an orbital period ($P$) shorter than 10 days. (b) \textbf{Warm Jupiters (WJ)} with 10 days$< P < 365$ days, (c) \textbf{Cold Jupiters (CJ)} as $P> 365$ days. In our sample, we have 59 hot Jupiters, 68 warm Jupiters, and 112 cold Jupiters. We investigated whether the host star metallicities ([Fe/H]) of these three groups show any similarities or differences.
In Figure \ref{fig:3}(a) we show the histograms of [Fe/H] distributions for hot, warm, and cold Jupiter hosts. We have fitted each normalized histogram with a Kernel Density Estimate and those are shown as black curves. The vertical black lines represent the medians of these distributions. Inspection of these distributions reveals that [Fe/H] distributions of WJ and CJ hosts are similar, but that of HJ hosts is different (See Figure \ref{fig:3}(a)). WJ and CJ hosts have a low metallicity tail, which is absent in the case of HJ hosts. The metallicity distribution of HJ hosts is flatter and has a smaller spread. The medians of WJ and CJ hosts are very close to each other, but the median [Fe/H] of HJ hosts is higher. 

In Figure \ref{fig:3}(b) metallicity distribution of hot, warm, and cold Jupiter hosts are shown as violin plots. As can be seen, the hot Jupiter hosts are relatively metal-rich while the warm and cold Jupiter hosts, on the other hand,  have low metallicity tails and have a similar median metallicity.

% we have shown all the planets as colored points in the orbital period and host star [Fe/H] plane. The black dashed lines separate HJ, WJ, and CJ. The big circles and associated errorbars represent the median and Median Absolute Deviation (MAD) of three distributions, respectively. Qualitatively, all three categories of Jupiter hosts have somewhat similar MAD, but median metallicity of HJ hosts is significantly higher than the WJ and CJ hosts. 
In Figure \ref{fig:3}(c) we have shown the cumulative distributions of [Fe/H] of three types of Jupiter hosts. The cumulative distributions also show HJ hosts are distinctly more metal-rich than WJ and CJ hosts. As one can see from Figure \ref{fig:3}(c), $\sim40\%$ of WJ and CJ hosts have subsolar metallicity ([Fe/H]$<0$) while that is true for only $\sim20\%$ of HJ hosts. 
\begin{figure}[htpb!]
    \centering
    \includegraphics[width=0.47\textwidth]{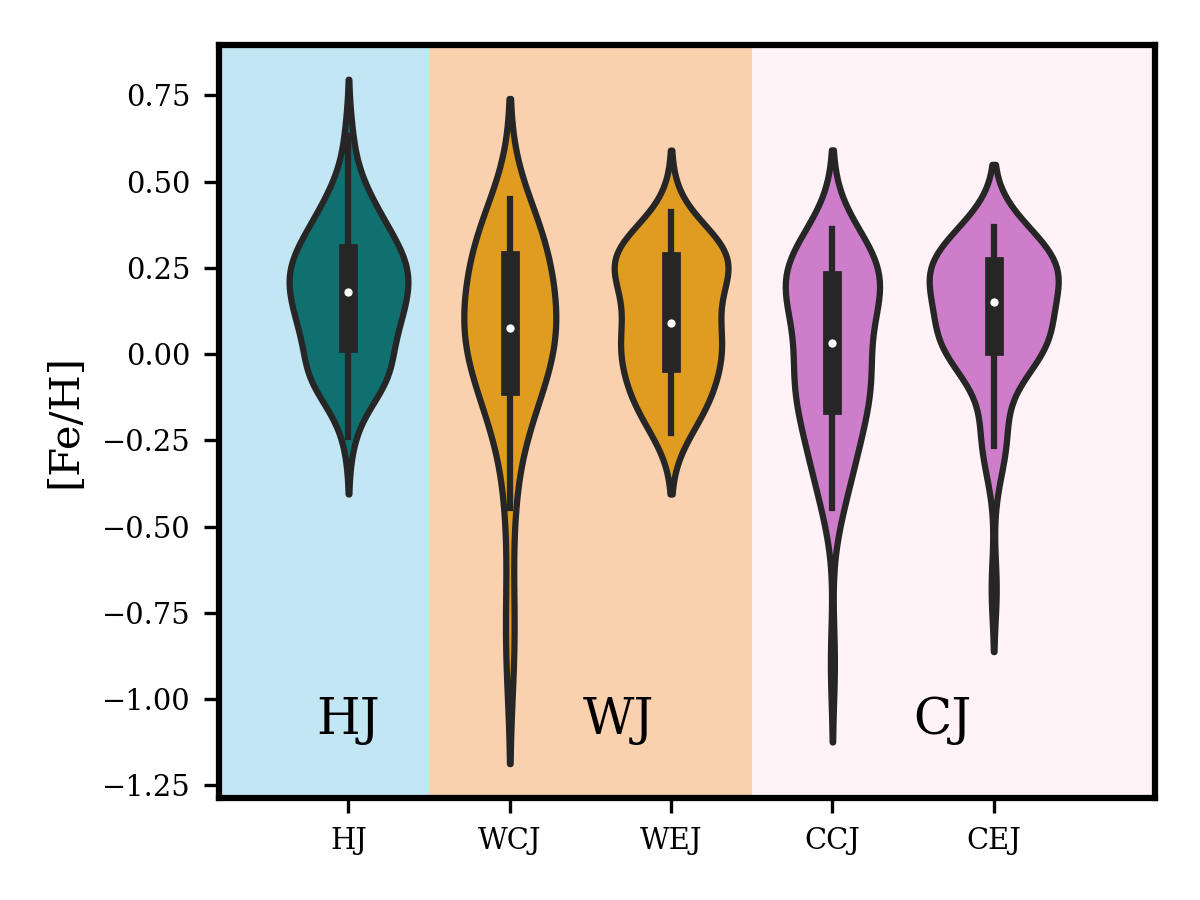}
   \caption{Host star [Fe/H] for high and low eccentricity planets among HJ, WJ, and CJ. For each group, the low eccentric or ``circular" subgroup is shown on the left, and eccentric subgroup is shown on the right. HJ is not subdivided into two subgroups due to lack of eccentric HJs in our sample. We note that,  only for cold Jupiters, the difference of the median host star [Fe/H] between eccentric and circular subgroups is greater than 0.1 dex. }
   \label{fig:all3ecc}
\end{figure}
To characterize these distributions quantitatively, we computed and compared the following quantities:  Median as a measure of the central tendency, Median Absolute Deviation (MAD) as a measure of the dispersion, Kurtosis, to characterize the flatness of the distribution, and Skewness, to characterize asymmetry and longer tail compared to a normal distribution towards a particular direction. For all the groups within our sample, these quantities are computed and summarized in Table \ref{tab:scores}.

We find that, if we only divide Jupiters into these three bins based on orbital period, HJ hosts seem to be relatively metal-rich than WJ and CJ hosts (similar to what has been reported in the literature by e.g., \citet{Narang18, CKS4}). Median metallicities ( Vertical black lines in Figure \ref{fig:3}(a)) of WJ and CJ hosts are similar to each other, and lower than that of HJ (See Table \ref{tab:scores}). We also see that metallicity distributions of HJ hosts are flatter and have a smaller spread, about a high metallicity value (0.18 dex). The spread is also almost symmetric about the median. This is reflected in the small values of Kurtosis and Skewness ( See Table \ref{tab:scores}). However, in the case of WJ and CJ hosts, the distributions have low metallicity tails, and they are asymmetric about the median. This is reflected in positive Kurtosis and negative Skewness scores as shown in Table \ref{tab:scores}.
%Observing three distributions, one may find that HJs have a metal-rich tail while WJ and CJs have a long tail towards low-metallicity (Figure \ref{fig:3}a). 
%The median metallicities of host stars of HJ, WJ, and CJ are Med([Fe/H])$_{\text{HJ}}= 0.18$, Med([Fe/H])$_{\text{WJ}}=0.08$ and Med([Fe/H])$_{\text{CJ}}=0.09$ respectively, i.e. HJ hosts are metal-rich on average, and no significant difference is seen between WJ and CJ hosts. All three distributions have quite a large Median Absolute Deviations (MAD); $\text{MAD}_\text{HJ}=0.13$, $\text{MAD}_\text{WJ}=0.18$, $\text{MAD}_\text{CJ}=0.15$.

\begin{table}[!h]
    \centering
    \begin{tabular}{|c|c|c|c|c|}
    \hline
    Sample & Median & MAD & Kurtosis & Skewness \\
    \hline 
    HJ & {0.18} & {0.13} & {-0.51} & {0.08} \\
    \hline
    WJ & {0.08} & {0.18} & {3.14} & {-1.38} \\
    \hline 
    CJ & {0.09} & {0.16} & {1.71} & {-1.15} \\
    \hline 
    CEJ & {0.15} & {0.12} & {2.66} & {-1.37} \\
    \hline 
    CCJ & {0.03} & {0.18} & {1.06} & {-0.92} \\
    \hline 
    \end{tabular}
    \caption{Statistical properties of the distribution of host star [Fe/H] for different classes of Jupiters}
    \label{tab:scores}
\end{table}
\begin{figure*}[htbp]
    \centering
    \includegraphics[width=\linewidth]{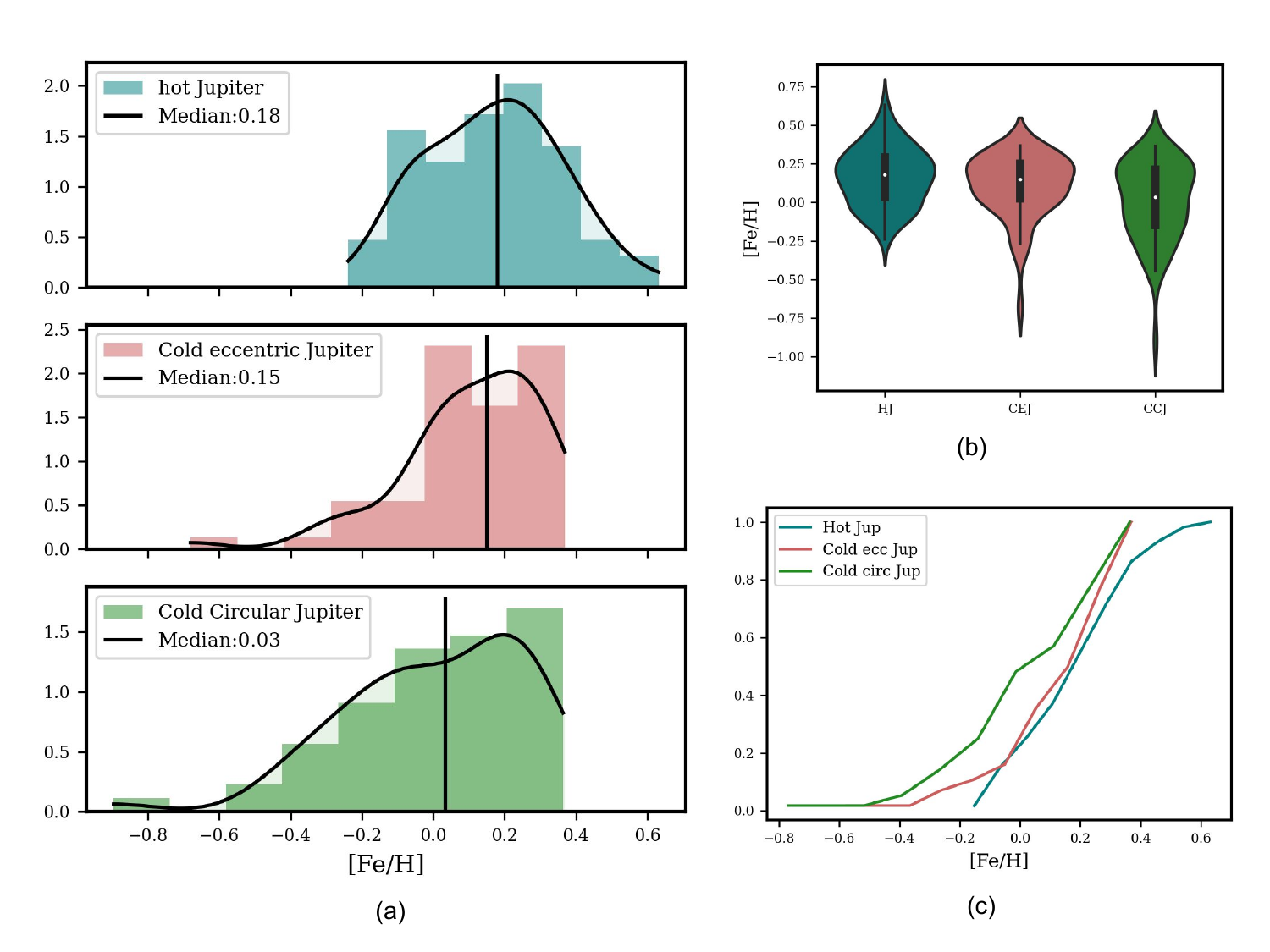}
    \caption{ (a)[Fe/H] distributions of HJ, CEJ, and CCJ hosts. The black curved lines are the Kernel Density Estimates from the colored histograms. The medians [Fe/H] distributions of HJ and CEJ are similar, while CCJ hosts have a lower median [Fe/H]. We also note that for HJ hosts, [Fe/H] distribution has a longer tail towards high [Fe/H], and CCJ has a long tail towards low metallicity. Also, the [Fe/H] distribution for CCJ is much flatter. (b) Violin plots showing [Fe/H] distributions of HJ, CCJ, and CEJ hosts. CCJs and CEJs are denoted by green and red respectively. (c) Cumulative distributions of [Fe/H] for HJ, CEJ, and CCJ. We can see that the central part of the host star [Fe/H] distribution of HJ and CEJ resemble each other.}
    \label{fig:cejccj}
\end{figure*}

However, in addition to their orbital periods, these Jupiters also show diversity in orbital eccentricity. Many independent works on planetary dynamics suggest eccentricity ($e$) is a tracer of the dynamic history of planets \citep[e.g.,][]{Rasio1996, Ford2008, Chatterjee2008, Tuhin2023}. Typically, planets gain large eccentricity either by interacting with each other \citep[e.g.,][]{Kozai1962, Chatterjee2008} or with an external body \citep{Munoz2016, Shara2016}. In the first scenario, metal-rich disks are preferred, while in the latter no such dependence on metallicity is expected \citep[e.g.,][]{Shara2016, Dawson18}. Therefore, it is interesting to investigate if host star metallicity is also correlated with the eccentricity of accompanying planets. With this in mind, we further subdivided HJ, WJ, and CJ into two bins with high and low orbital eccentricity. We chose $e=0.2$ as the dividing criterion between two bins. Hereafter, We will call the Jupiters with $e<0.2$ as the ones in "circular" orbits, and those with $e>0.2$ as planets in "eccentric" orbits. 
We find 56 cold Jupiters in "eccentric" orbits, and 56 cold Jupiters in "circular" orbits. Hereafter we will call them CEJ (Cold Jupiters in Eccentric orbits) and CCJ (Cold Jupiters in Circular orbits) respectively.
Warm Jupiters also show a smaller but significant variation in eccentricity. We find 33 warm Jupiters in eccentric orbits and 35 warm Jupiters in circular orbits. On the other hand, most of the hot Jupiters are in circular orbits. We find only 2 HJs in eccentric orbits.

In Figure \ref{fig:all3ecc}  host star [Fe/H] distributions for low and high eccentricity subgroups of warm (WJ), and cold Jupiters (CJ) are shown as violin plots. We don't show the eccentric subgroups of HJ, because only 2 HJs have high-eccentricity in our sample. The color palette is the same as in Figure \ref{fig:3}. The low eccentric or ``circular" subgroup is shown on the left and the eccentric subgroup is shown on the right. We find that the difference of median host star [Fe/H] between the eccentric and circular subgroups is greater than 0.1 dex, only for cold Jupiters. 

Therefore, for the rest of the paper, we treat CEJ and CCJ as two separate groups but combine eccentric and circular subgroups for HJ and WJs.
%The orbital period of planets is on the x-axis and the corresponding host star [Fe/H] is on the y-axis. 
%The vertical, dashed black lines represent the boundaries between hot, warm and cold Jupiters. Planets with $e>0.2$ are shown with lower triangle ($\nabla$) symbol and planets with $e<0.2$ are shown as circles ($\circ$). Different colors; i.e. blue, orange, and magenta are for HJ, WJ, and CJ respectively. The x and y coordinates of the big circles with black borders are the median orbital period and median [Fe/H] in each bin for Jupiters in low eccentric orbit, and the horizontal and vertical errorbars represent the corresponding MAD. To denote median period and median host star metallicity of the eccentric Jupiters, we use big lower triangle with black borders.  We find that the difference of median [Fe/H] of host stars of ``circular" and ``eccentric" planets is $>0.1$ only for Cold Jupiters. 

%With this in mind, we subdivided cold Jupiters into two eccentricity bins \citep[also see][]{Buchhave18}. We chose $e=0.2$ as the dividing criterion between two bins, as that makes the number of planets in both bins equal (56 each). 

In Figure \ref{fig:cejccj} we have shown the [Fe/H] distributions of CEJ and CCJ hosts along with HJ hosts. Figure \ref{fig:cejccj}(a) shows the three histograms and Kernel Density Estimations, and the vertical line corresponds to the median of each. Figure \ref{fig:cejccj}(b) is a violin plot of the three distributions. It is evident from the figure that host stars of HJ and CEJ are more metal-rich than CCJ hosts. 
We find that Cold ``eccentric" Jupiters (CEJ) and Cold ``circular" Jupiters (CCJ) have different distributions of [Fe/H], with eccentric cold Jupiters being relatively more metal-rich on average (median [Fe/H]$= 0.15$). The average host star metallicity of cold circular Jupiters, however, is close to the solar value (median [Fe/H]$= 0.03$) (See Table \ref{tab:scores}). The distribution of [Fe/H] also seems to be flatter in the case of CCJ. We note that with a larger sample and homogeneous dataset our result qualitatively follows the findings of \cite{Buchhave18}.

%To examine if eccentric warm Jupiters also have a preference around metal-rich host stars, we subdivided all planets into high and low eccentricity bins. 

Comparing [Fe/H] distributions of HJ, WJ, CEJ, and CCJ we arrive at the following conclusions:

\begin{enumerate}
    \item The median difference between [Fe/H] distributions of HJ and CJ hosts is 0.09, and between HJ and WJ hosts, it is 0.1. On the other hand, the median difference between WJ and CJ hosts is only -0.01. 

    \item  Distributions of [Fe/H] among CJ hosts show a dichotomy, dependent on the eccentricity of the accompanying Jupiter. The median difference between CEJ and CCJ hosts is 0.12. The median difference of [Fe/H] distributions between HJ hosts and CCJ hosts is 0.15. The median [Fe/H] of HJ and CEJ hosts are very similar, with a difference of only 0.03.

    \item  Based on the median differences and shapes of the distributions, [Fe/H] distributions of CCJ hosts differ from HJ hosts and CEJ hosts. On the other hand, [Fe/H] distributions of HJ and CEJ hosts are very similar. It hints towards the possibility that Cold Eccentric Jupiter hosts and Hot Jupiter hosts belong to the same underlying population, and Cold Circular Jupiter hosts are from a different population.

\end{enumerate}
 \begin{figure*}[htbp]
    \centering
    \includegraphics[width=\textwidth]{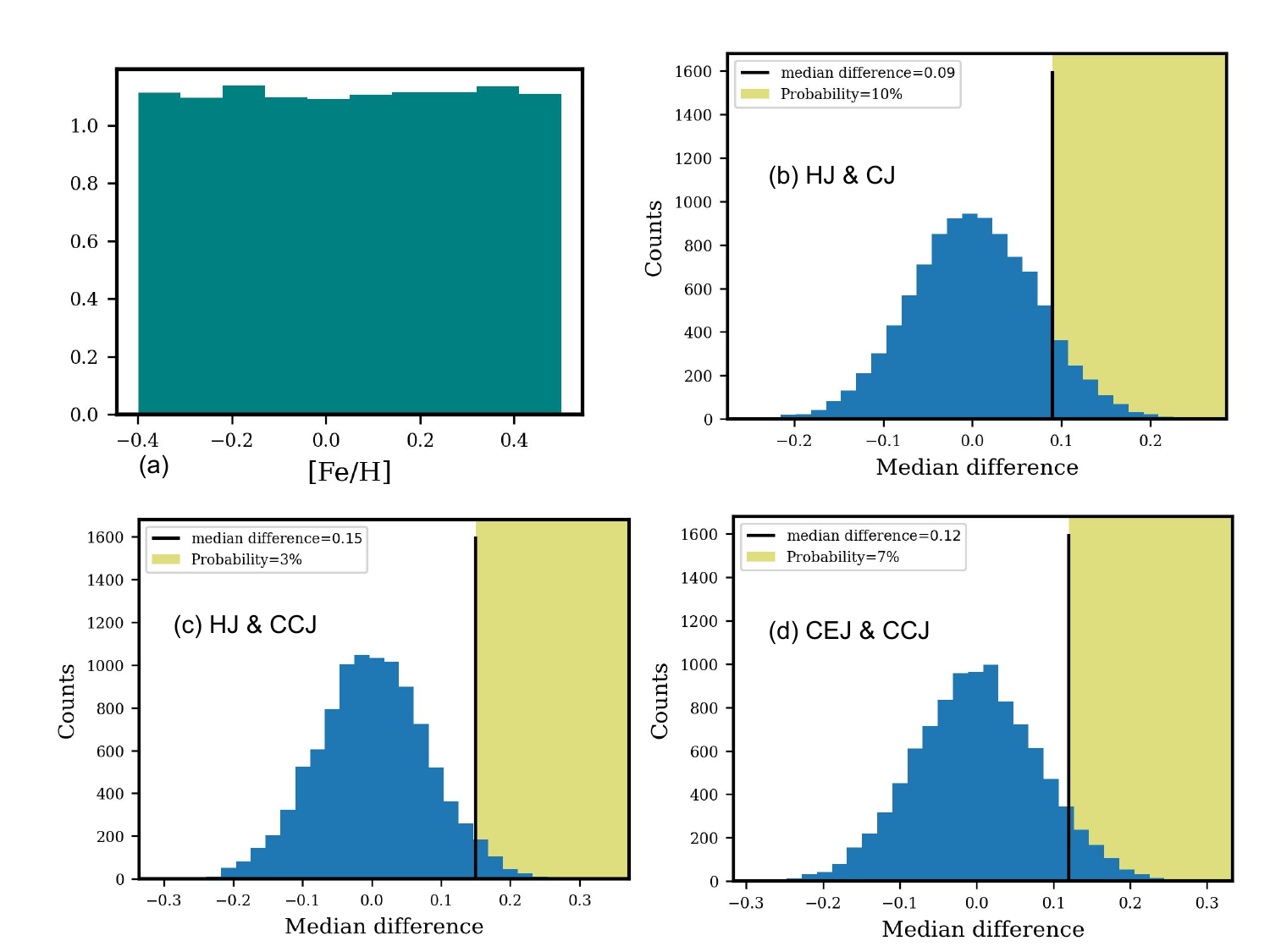}
    \caption{Monte Carlo Analysis 1: Can random sampling from an underlying \textbf{uniform distribution} of host star [Fe/H] produce the observed results? (a) Shows histogram of the uniform sample. The blue histograms are the distribution of median differences for (b) HJ and CJ (c) HJ and CCJ and (d) CEJ and CCJ. The vertical black solid lines in all these figures represent the observed median differences. The region right to this black solid line is colored yellow, here the median difference $\geq$ observed median difference. The probability of median difference $\geq$ observed median difference is the area under the histogram in the yellow shaded region divided by the total number of counts. )}
    \label{fig:monte_uni}
\end{figure*}
The immediate questions that arise from these results are the following. 
\begin{enumerate}
    \item How statistically significant are these results? 
    \item Are these results coming from the true nature of these distributions, or is it due to a random sampling of points from a different underlying distribution?
    \item Do other properties of host stars (e.g., age) of the sample also follow the similarities and differences between HJ, CEJ, and CCJ populations? 
    \item What formation and evolution channels of Jupiter formation do these results support? 
    \item Do these results suffer from selection effects and observational biases?
\end{enumerate}

We address these questions in the following Sections; we discuss questions 1 \& 2 in Section \ref{sec: stat}, question 3 in Section \ref{sec: age}, and questions 4 \& 5 in Section \ref{sec:dis}. 

\section{Statistical analysis of the results}\label{sec: stat}

%Are the Jupiters at various orbital distances with various orbital eccentricities come from the same population? How different are the distributions of their host star metallicities? Is the observed difference a result of random sampling? To have a quantitative answer to these questions, we have followed a systematic procedure. We note that the distributions that we are comparing are only limited samples, and are not necessarily representative of the original distributions.  

%We find that medians of the [Fe/H] distribution of HJ and CEJ hosts differ from the CCJ hosts by greater than 0.1 dex. On the other hand, the median [Fe/H] of HJ hosts and CEJ hosts are very similar, with a difference of 0.03 dex. Histograms and cumulative distributions of [Fe/H] of HJ and CEJ hosts appear similar, but both appear different from CCJ hosts (See Section \ref{sec:monte} ). These findings hint at the possibility that the HJ and CEJ hosts belong to the same underlying population while CCJ hosts are characteristically different. 

In this section, several tests to compute the statistical significance of these results are discussed. %The major testing is about whether two populations of Jupiters are in the same population or not, and for that, we need to te

First, we approach the problem using Monte Carlo analysis. We assume all Jupiter hosts belong to the same underlying population, i.e. they have one common [Fe/H] distribution for all Jupiter hosts, HJ, WJ, and CJ alike. Then we randomly draw samples from this distribution, divide them into separate groups, and label them as HJ, CJ, CEJ, or CCJ. We keep the sample size of each group equal to the observed sample size of that group. Then we compute the probability of finding the difference of the medians of the two groups greater or equal to the observed median differences. (See Section \ref{sec:monte}). If this probability is greater than $5\%$, we do not rule out the possibility that our observed median difference is an outcome of random draws of subsamples from a common metallicity distribution of host stars. Otherwise, we rule out the possibility. 

In section \ref{sec:statest}, we will perform several non-parametric null hypothesis tests between the observed [Fe/H] distributions of HJ, CEJ and CCJ hosts. 
% However, for Monte Carlo analysis, we have to assume the shape of the underlying metallicity distribution of host stars, and we do not have prior knowledge of the true distribution. To compare the observed samples directly in a quantitative way without assuming any form of the underlying distribution, we should use non-parametric statistical tests. 
 
% These tests compute the confidence level of rejecting the null hypothesis. Our null hypothesis is that all Jupiter hosts are from the same parent population, i.e. all Jupiter hosts share a common metallicity distribution. All [Fe/H] measurements are similar to draw random samples from this parent distribution of [Fe/H]. We will reject the null hypothesis only if our confidence level of rejection is $\geq 95\%$.
 
 %There are three ways to compare two distributions. (1) Comparison of the central tendency (2) Comparison of the tails (3) Comparison of the shape. 
 
 %However in our cases, we only have subsamples of the original distributions, and we do not have the prior information of the true distributions. Therefore, we perform a set of non-parametric, two-sided statistical tests with bootstrap resampling among the class of Jupiters in Section \ref{sec:statest}, and check if the rejection of the null hypothesis is possible.
 
\subsection{Monte Carlo Analysis}\label{sec:monte}

%If all Jupiter hosts belong to the same underlying population, then what is the probability for random sampling to produce our observed result? In other words, what is the probability that the true [Fe/H] distributions of HJ, CCJ and CEJ host stars are not different at all, and only the random sampling of limited size gives rise to the observed differences? We examine this in the following.

%In Figure \ref{fig:monte_uni}(a) we show the assumed uniform distribution of [Fe/H] of Jupiter hosts. In other subfigures, we show the distributions of median differences we get after 10000 random draws as blue histograms. The yellow-shaded region represents where the median difference $\geq$ observed median difference. The area under the blue histograms in this region,  divided by total counts represents the probability of occurrence of a median difference $\geq$ observed difference.

%We start by assuming an underlying distribution of host star metallicities. In our sample of Jupiters, we have 59 HJ, and 112 CJ. Among 112 CJ we have 56 CEJ and 56 CCJ. We draw random points of the exact sizes of the observed sample from the underlying distribution of the host star [Fe/H], compute the medians of the drawn sample, and obtain the difference of these medians. We repeat this exercise 10,000 times. As a result, we get a distribution of the median difference between the two groups drawn from the same distribution. We obtain the probability of getting a median difference from this distribution to be greater than the observed median difference. 
\begin{figure*}[htbp]
    \centering
    \includegraphics[width=\textwidth]{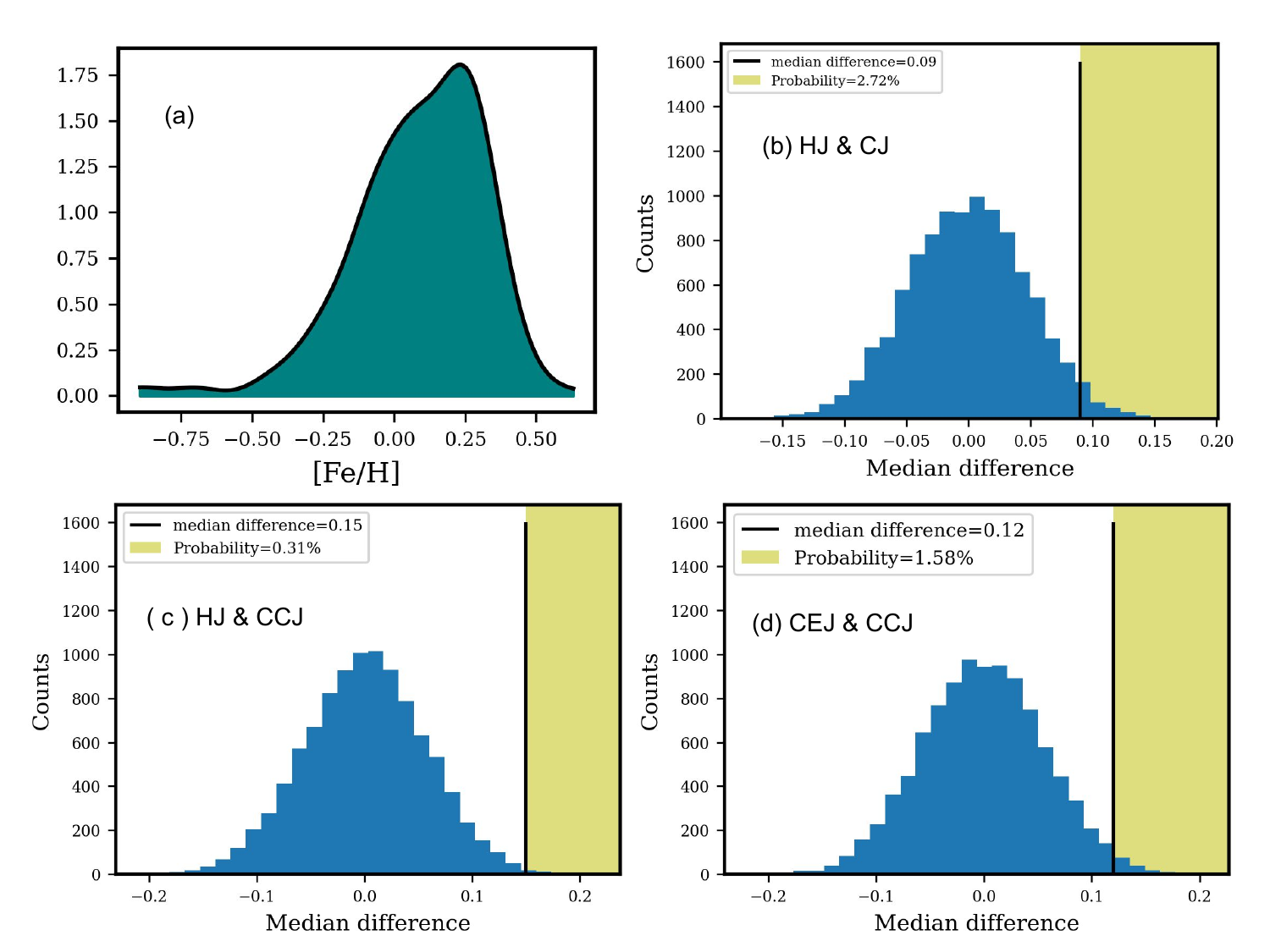}
    \caption{Monte Carlo Analysis 2:  Can random sampling from an underlying distribution of host star [Fe/H] produce the observed results? (a) The underlying distribution is Kernel Density Estimation to the [Fe/H] distribution of Jupiter hosts in our sample. We draw random samples from this distribution repeatedly of the exact sizes of our observed sample and compute the median difference. The blue histograms are the distribution of median differences for (b) HJ and CJ (c) HJ and CCJ and (d) CEJ and CCJ. The vertical black solid lines in all these figures represent the observed median differences. The region right to this black solid line is colored yellow, here the median difference is greater than the observed median difference. The probability of median difference $\geq$ observed median difference is the area under the histogram in the yellow shaded region divided by the total number of counts. )}
    \label{fig:monte_real}
\end{figure*}

First, we assume the underlying distribution of host star [Fe/H] is uniform (See Figure \ref{fig:monte_uni} (a)). We note that $\geq95\%$ of all Jupiter hosts have [Fe/H] values between -0.4 to 0.5. We assume all Jupiter hosts have an equal probability of having a [Fe/H] value anywhere in this range. The median difference of host star [Fe/H] between HJ and CJ from our observed sample is 0.09 dex. We randomly draw two samples of the exact sizes of HJ \& CJ in observed samples from this uniform distribution and compute the median difference between the two samples. This exercise is repeated 10,000 times, so we obtain a distribution of median differences of two subsamples drawn from the same underlying distribution. In Figure \ref{fig:monte_uni}(b) we show this distribution as the blue histogram. The vertical black line in Figure \ref{fig:monte_uni}(b) represents the observed median difference of host star [Fe/H] between HJ and CJ (0.09 in this case). The area right to this vertical black line is shaded yellow. The obtained median difference is greater than the observed median difference in this region.
We find that the probability of occurrence of the median difference $\geq$ observed difference by simply taking the ratio of the number of occurrences where the obtained median difference is larger than the observed median difference, and the total number of draws (10000). For HJ and CJ, the probability is $10\%$ (See Figure \ref{fig:monte_uni} (b)). 

In the case of CCJ and CEJ the sample size is 56 for both. The observed median difference of host star [Fe/H] distributions between HJ and CCJ is 0.15 and between CEJ and CCJ is 0.13. Following the same procedure, we compute the probability that a random draw from the underlying uniform distribution produces this result or a larger median difference. Between HJ and CCJ, the probability is only $3\%$, and between CEJ and CCJ the probability is $7\%$. 

Therefore, if [Fe/H] values of all Jupiter hosts belong to the same uniform distribution, the probability of getting the observed median differences or greater are (1) $10\%$ for HJ \& CJ (2) $3\%$ for HJ and CCJ (3) $7\%$ for CEJ and CCJ. 

Therefore, we can only rule out the possibility of having a common parent population of metallicities, only for HJ and CCJ, according to the threshold we have set. 
%if drawn from the same underlying uniform distribution there are $90\%$, $97\%$ and $93\%$ chance of getting a median difference less than the observed values respectively for the three cases.  

However, the underlying distribution of [Fe/H] of Jupiter hosts is very unlikely to be uniform. As multiple works suggest, Jupiter hosts are preferentially metal-rich \citep[e.g.,][]{Santos17, Mulders2018, Narang18, ZhuandDong2021}. Therefore, to represent a more realistic scenario, we take the Kernel Density Estimate (KDE) of the [Fe/H] distribution of our observed sample to be the underlying distribution for all Jupiter hosts. Then we repeat the same exercises as above. In Figure \ref{fig:monte_real} we summarize our findings.

Figure \ref{fig:monte_real}(a) shows the KDE of the [Fe/H] of all Jupiter hosts in our sample. We assume this to be the underlying distribution for all Jupiters. First, we draw samples of the exact sizes as HJ and CJ samples. Figure \ref{fig:monte_real}(b) shows the resulting distribution of the median difference between HJ and CJ. We find that the probability of the median difference being greater than the observed is $2.72\%$. In the case of HJ and CCJ ( See Figure \ref{fig:monte_real}(c)) this probability is only $0.31\%$. Finally, between CEJ and CCJ the probability is $1.58\%$ (See Figure \ref{fig:monte_real}(d)).

Therefore, if [Fe/H] values of all Jupiter hosts belong to the same distribution as approximated by KDE of our sample, the probability of getting the observed median differences or greater are (1) $2.72\%$ for HJ \& CJ (2) $0.31\%$ for HJ and CCJ (3) $1.58\%$ for CEJ and CCJ. 

From the preceding analysis, we can safely conclude that 
\begin{enumerate}
    \item If we assume a common, uniform [Fe/H] distribution of the Jupiter hosting stars, and [Fe/H] distributions of HJ,CCJ and CEJ are randomly drawn subsamples from this common parent distribution, we find the probability of getting the observed or a greater median difference between them is $<5$\% only for HJ and CCJ hosts. Therefore we can rule out the possibility that HJ and CCJ hosts have a common uniform parent [Fe/H] distribution with $>95$\% confidence. 
%we can reject the null hypothesis that [Fe/H] samples HJ and CCJ hosts come from the same distribution with $>95\%$ confidence. But we can't reject the null hypothesis either for CEJ and CCJ hosts or for HJ and CJ hosts.

    \item If we assume the [Fe/H] distribution of the Jupiter host stars follows the KDE of [Fe/H] our sample then we can rule out the possibility that HJ and CCJ belong to the same parent population with $99.99\%$ confidence. We can do the same for HJ and CJ hosts with $97.28\%$ confidence and CEJ and CCJ hosts with $98.42\%$ confidence. It hints strongly that CEJ and HJ hosts are well-separated populations from CCJ hosts, in terms of metallicity.
\end{enumerate}

\subsection{Statistical Tests}\label{sec:statest}
% we use the two-sample Kolomogrov-Smironov (KS) test between metallicity distributions. The results are summarised in Table \ref{tab:ks}. We note that the pvalue we find using a much larger and homogeneous sample is less statistically significant than \cite{Buchhave18}. Our key results are: (1) Hot Jupiters and Cold Jupiters in circular orbits belong to two different populations. We can reject the null hypothesis with $\sim 97$ \% confidence. (2) Cold circular Jupiters and Cold eccentric Jupiters also belong to two different populations, with $\sim$ 96\% certainty. (3) Warm Jupiters can not be distinguished from any of the other populations.%

We have found, in Section \ref{sec: corr} that \textit{Gaia} DR3 metallicities ([Fe/H]) of host stars of HJ and CEJ show a similar distribution, however, these distributions differ from the [Fe/H] distribution of CCJ hosts. This indicates perhaps HJ and CEJ hosts belong to the same underlying population, whereas CCJ hosts come from a different population. Now we will use various statistical tests to compare the central tendencies, dispersion, and tails of these distributions, and determine the statistical significance of these results. 

Since we do not have any prior information about the true metallicity distribution of these stars, to compare the samples we use non-parametric, distribution-free tests \citep[e.g.,][]{nonpar}. All these tests check if the null hypothesis is true, i.e. if the two populations underlying the two samples are identical. The alternative hypothesis is that they are not identical. 

In these tests, the first step is to compute the test statistic, which quantifies the difference between two or more data groups (Datasets of [Fe/H] of HJ, CJ, CEJ, and CCJ hosts in our case). Then, we compute the p-value, which represents the probability of observing a test statistic as extreme as, or more extreme than, the one computed from the data, assuming that the null hypothesis is true (i.e., there is no real difference between the data groups). A smaller p-value suggests stronger evidence against the null hypothesis, indicating that the observed data is unlikely to have occurred by random chance alone.

To reject the null hypothesis, we set the confidence level to 95\% or the threshold p-value to 0.05. If the resultant p-value of all the tests for two samples is $\leq 0.05$ we reject the null hypothesis with $\geq 95\%$ confidence.
\begin{table}[!h]
    \centering
    \begin{tabular}{|c|c|c|c|c|}
    \hline
         Test & Sample1 & Sample2 & Statistic & p-value  \\
         \hline 
         MW U  & HJ & CJ & 4063.5 $\pm$ 200.5 & 0.014 $\pm$ 0.013 \\
         {Test} & HJ & CCJ & 2222.0 $\pm$ 111.0 & 0.001 $\pm$ 0.001 \\
         {} & CEJ & CCJ & 1944.0 $\pm$ 112.5 & 0.025 $\pm$ 0.024 \\
         {} & HJ & CEJ & 1846.5 $\pm$ 118.5 & 0.264 $\pm$ 0.217 \\
         \hline 
         KW  & HJ & CJ & 6.09 $\pm$ 3.12 & 0.014 $\pm$ 0.013 \\
         {Test} & HJ & CCJ & 10.18 $\pm$ 3.92 & 0.001 $\pm$ 0.001 \\
         { } & CEJ & CCJ & 4.79 $\pm$ 2.77 & 0.024 $\pm$ 0.023 \\
         {} & HJ & CEJ & 1.253 $\pm$ 1.108 & 0.263 $\pm$ 0.215 \\
         \hline 
         KS  & HJ & CJ & 0.22 $\pm$ 0.04 & 0.03 $\pm$ 0.03 \\ 
         {Test} & HJ & CCJ & 0.32 $\pm$ 0.05 & 0.004 $\pm$ 0.004 \\ 
         { }  & CEJ & CCJ & 0.32 $\pm$ 0.05 & 0.006 $\pm$ 0.006 \\
         {} & HJ & CEJ & 0.198 $\pm$ 0.03 & 0.175 $\pm$ 0.132 \\
         \hline 
    \end{tabular}
    \caption{Results of two-sample, two-sided statistical tests (a) Mann-Whitney (MW) Test (b) Kruskal-Wallis (KW) Test (c) Kolmogrov-Smironov (KS) Test. Note: We must have a p-value $<0.05$ to reject the null hypothesis. The standard scipy packages have been used to obtain the test results. }
    \label{tab:tests}
\end{table}

We do not know if the observed samples are the true representatives of the underlying distributions. Therefore, instead of comparing the observed datasets directly, we use bootstrapped resampling, i.e., we repeatedly draw random samples from the datasets with replacements. %If a dataset has N data points, we randomly draw N points from it, with replacements, and call it a bootstrapped sample.
We compute the statistical tests between the bootstrapped samples in each iteration and finally get distributions of test statistics and p-values. We obtain the median and MAD of these distributions, and these results are summarised in Table \ref{tab:tests}.

To compare the central tendencies and dispersion of two distributions, we use the Mann-Whitney (MW) U test \citep[e.g.][]{MannWhitney47} and the Kruskal-Wallis (KW) test \citep[e.g.,][]{kruskalWallis52}. For comparison of the cumulative distributions, we use the Kolmogrov-Smironov (KS) test \citep[e.g.,][]{Smironov1948, Hodges1958}. A short description of these tests can be found in the Appendix \ref{sec: appen}.

%To reduce the uncertainty, and to remove the effect of possible outliers, we use bootstrap resampling for each test \citep[e.g.,][]{FeigelsonBabu}. If a dataset has N number of points, we randomly draw N points from the dataset, allowing replacements. This set of N points is a bootstrapped sample of the original dataset.   

%and compute these statistical tests for each of the bootstrapped samples. Therefore we get distributions of test statistics and p-values. 

The comparison between host star [Fe/H] distributions of HJ, CJ, CCJ, and CEJ is compiled in Table \ref{tab:tests}. The null hypothesis in each case is the two samples (sample 1 \& 2 in Table \ref{tab:tests}) belong to the same parent population. If we look at the MW U test and KW test results, the dataset pair \{HJ, CJ\} has the p-value $<$ 0.05. However, the obtained p-value for the KS test statistic between the pair \{HJ, CJ\} is $0.03\pm 0.03$, so the upper limit is marginally greater than our threshold $0.05$. This implies the medians of HJ and CJ are well-separated and the difference is significant, but the difference between their cumulative distribution is not significant. 

For the dataset pairs \{HJ, CCJ\} and \{CEJ, CCJ\}, we have p-value$< 0.05$ in all cases.  
Therefore, for \{HJ, CCJ\} and \{CEJ, CCJ\} we can safely reject the null hypothesis and conclude they do not come from the same underlying population. 

However, for the pair \{HJ, CEJ\} we have the smallest test statistic for all the tests, and also a large p-value $>0.05$ corresponding to the test statistic. Therefore we can not reject the null hypothesis in this case. 

%We have set our threshold p-value to be 0.05 to reject the null hypothesis. We see that all of the pairs (HJ, CJ), (HJ, CCJ), and (CEJ, CCJ) have a p-value $< 0.05$ with MW U test. 

%The KS test scores between different groups of Jupiters are listed in Table \ref{tab:tests}. We have set the minimum criteria to reject the null hypothesis by getting a p-value $<0.05$. From the Table \ref{tab:tests} we see that between HJ and CCJ, and CEJ and CCJ this criteria is satisfied, and therefore in these cases we can reject the null hypothesis and conclude, [Fe/H] distributions of hosts of HJ and CCJ , and CEJ and CCJ, are not likely to be from the same parent population. While we can not reject the null hypothesis while comparing HJ and CEJ. 

All of the above statistical test results strengthen our findings of section \ref{sec: corr} and we can safely conclude the following:

\begin{enumerate}
    \item Regarding [Fe/H] distributions, host stars of HJ and CCJ do not belong to the same parent population. Similarly, host stars of CEJ and CCJ belong to different parent populations. 

    \item In terms of [Fe/H] distributions, host stars of HJ and CEJ likely belong to the same parent population. There is no evidence in the data to suggest otherwise. 
\end{enumerate}

Therefore, our results so far, support the possibility that HJ hosts and CEJ hosts have similar properties, and CEJs might be the progenitors of HJs, both forming from disks with super-solar metallicity. On the other hand, CCJs represent a separate class and form in relatively lower metallicity disks (Also see \citet{Buchhave18}).

\section{Age-Orbital Period-Metallicity Correlation}\label{sec: age}
%\subsubsection{Planet multiplicity - metallicity connection}\label{multi}
%According to HEM scenario triggered by planetary perturber, we expect a correlation with metallicity and multiplicity. In our limited sample, if we compare the metallicities of single and multiple Jupiter hosts, we find that on average, multiple Jupiter hosts have a slightly higher metallicity ([Fe/H]$=0.18\pm 0.13$) than the single Jupiter hosts ([Fe/H]$= 0.08\pm 0.15$). However, the KS test can not separate out two populations, with a pvalue of 0.18. We note that this sample is not corrected for possible detection biases of multiplicity and a true correlation may only be found after such corrections.

 %I need help writing this section

 \begin{figure*}[htbp]
    \centering
    \includegraphics[width=\textwidth]{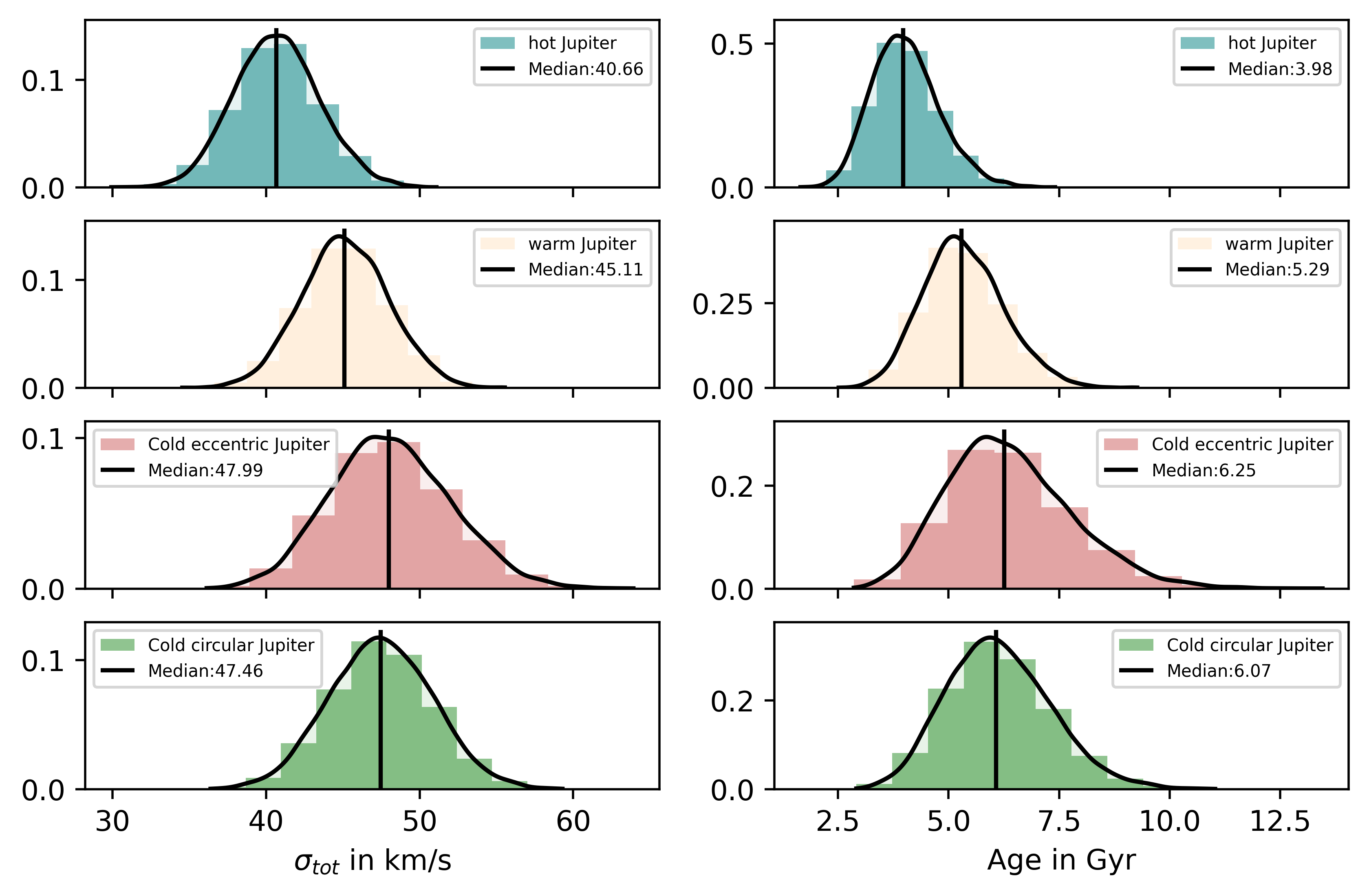}
    \caption{Velocity Dispersion (left column) and Age distributions (right column) of HJ, WJ, CEJ and CCJ hosts. The distributions are obtained using the bootstrapped resampling method described in the text. We find that hot Jupiter hosts are younger with the smallest velocity dispersion. }
    \label{fig: age-vel}
\end{figure*}
As the universe evolves, ISM gets enriched in metals. Therefore stars formed at a later epoch, i.e. younger stars, have a higher metallicity compared to the older stars, on average. This results in a positive correlation between stellar metallicity and age over Galactic timescales \citep[e.g.,][]{Carlberg1985, Meusinger1991,nordstrom2004}. On the other hand, as metallicity increases in the molecular cloud, the efficiency in giant planet formation also increases\citep[e.g.,][]{IdaLin2005, WyattClarke2007, Mordasini12, Piso2015}. It has been adequately demonstrated in the literature that giant planet host stars are metal-rich compared to field stars and small planet hosts \citep[e.g.,][]{Gonzalez1997, Santos2003, Fischer2005, Santos2006, Urdy_Santos2007, Narang18}. One might also expect these stars to be relatively young, from the metallicity-age correlation of the stars, and some works in the literature suggest the same \citep[e.g.,][\color{bibblue}{Narang et al. (2024), Under review}]{Swastik2022}. Several independent works have found HJ hosts to be younger on average than the field stars \citep[e.g.,][]{Mustill2022, Blaylock2023, Hamer2019}. However, the reason for the relative age difference is debated, as it might be a consequence of the age-metallicity correlation \citep[e.g.,][\color{bibblue}{Narang et al. (2024), Under review}]{Swastik21, Swastik2022, Narangt}, or the destruction of the older population of HJs by their host stars \citep[e.g.,][]{Hamer2019}, or a combination of both. In this work, we are comparing host stars of Jupiters only, but these Jupiters are located at various orbital distances. In the previous sections, we have discussed their differences in metallicity. Now we ask, do these Jupiter hosts follow a similar trend in terms of age?   

However, it is non-trivial to measure the individual ages of main-sequence stars, especially for solar or lower mass stars \citep[e.g.,][]{Soderblom2010}. Methods of obtaining ages of individual stars include isochrone-fitting \citep[e.g.,][]{Soderblom2010,Takeda2007}, asteroseismology measurement \citep[e.g.,][]{Bazot2008}, using empirical relation from stellar spindown \citep[e.g.,][]{barnes2009}. However, isochrone-fitted age has large uncertainties \citep[e.g.,][]{Takeda2007}. Ages determined by modeling asteroseismology measurements have much less uncertainty, but these observations are resource-intensive, and a homogeneous asteroseismology measurement for our sample is not available yet \citep[e.g.,][]{Bazot2008}. Similarly, a homogeneous dataset for the stellar rotation period is also not available for the stars in our sample.
%One of the recommended methods is to measure the dispersion of the galactic velocity for a bunch of stars and convert that to ages.
Nevertheless, we are not interested in the ages of individual stars, rather we want to know how the groups of HJ,WJ, CEJ and CCJ host stars differ in terms of their average age.
One of the ways to obtain an ensemble age for a group of stars is to use the dispersion in their Galactic space velocities as a proxy for age. This has been well-established and widely used in the literature \citep[e.g.,][]{Wielen1977,Carlberg1985, Meusinger1991, Dehnen_Binney1998, Binney2000, Manoj2005,AumerBinney2009, Schronrich2010, Sharma2014, YuLiu2018}. Using this method, \textcolor{bibblue}{Narang et al. (2024), (Under review)} have shown, that hosts of Jupiters are metal-rich and younger than the field stars and small-planet hosts. Similar findings, using different methods are reported in \citep[e.g.,][]{Swastik2022, Miyazaki2023}{}{}.
%In the context of galactic chemical evolution, high-mass planetary systems are found to be younger, as their hosts are relatively rich in iron peak elements, compared to low-mass planets and field stars \citep{Swastik2022} \textcolor{blue}{Narang et al 2023, (\textit{under review})} also find that hot Jupiter hosts are relatively metal-rich and younger than warm and cold Jupiter hosts. \textit{Gaia}DR3 provides velocity information for (number) stars in the solar neighborhood. 
We applied a similar analysis to our sample of stars hosting hot, warm, and cold Jupiters in circular and eccentric orbits. 

Gaia DR3 provides parallax $(\pi)$, proper motion in right ascension $(\text{pmRA}, \mu_\alpha)$ and declination $(\text{pmDE},\mu_\delta)$ and radial velocity (RV, $\gamma$) information for all the stars in our sample. First, we calculate the space velocity components $\left(U,V,W\right)$ from the observed radial velocity and proper motion. 
\begin{align}\label{eq: veleq}
    \begin{bmatrix}
        U \\ 
        V \\
        W 
    \end{bmatrix} &= \textbf{B}\begin{bmatrix}
        \gamma \\ \frac{k \mu_\alpha }{\pi} \\ \frac{k\mu_\delta}{\pi}
    \end{bmatrix}
\end{align}
Where $k$ is constant, $k=4.740470$ $\text{km s}^{-1}$  \citep[e.g.,][]{Narang22,2020AJ....159..166U}. \textbf{B} is a $3\times 3$ matrix that transforms the equatorial velocity vector to the Cartesian system with the z-axis pointing towards the Galactic North Pole. The above equation \ref{eq: veleq} gives us heliocentric space velocities. Subtracting the solar velocity gives $(U_\odot, V_\odot, W_\odot)=(11.1,12.24,7.25)$ $\text{km s}^{-1}$ \citep{Schronrich2010} gives us space velocities with respect to our Local Standard of Rest. 

Now, for each of the ensembles of Hot, Warm, and Cold Jupiter hosts we compute the dispersion in galactic velocity as the sum of the dispersions in individual components:

\begin{equation}\label{eq: sigma}
    \sigma_{\text{tot}}=\sqrt{\sigma_U^2+\sigma_V^2+\sigma_W^2}
\end{equation}
Where, $\sigma_U^2=\dfrac{1}{N}\sum_{i=0}^N \left(U_i-\Bar{U}\right)^2$

In this way, we get the one value of velocity dispersion for each bin (HJ, WJ, CJ, CEJ, CCJ). However, there are two major sources of errors in this computed dispersion: 
\begin{enumerate}
    \item The errors in the measured values of proper motions and radial velocities.
    \item Sampling error. Since our sample in each bin is only a subset of a larger population, the velocity dispersion we compute may not represent the true velocity dispersion of the population.
\end{enumerate}
 
For this reason, we compute a distribution of $\sigma_{tot}$ with bootstrap resampling. Suppose, in a bin, we have N number of host stars. We select N host stars randomly out of them with replacements, i.e. same star can be selected twice. Measurements of parallax, pmRA, pmDE, and radial velocities, are available for all these stars, and each measurement has an associated error. 
We consider each measurement to be a Gaussian with the mean ($\mu$) as the reported value, and the standard deviation ($\sigma$) as the associated error, and we draw random values out of these distributions. Using these values, following equations \ref{eq: veleq} and \ref{eq: sigma} we compute $\sigma_{tot}$. We repeat this exercise 10000 times to get a distribution of $\sigma_{tot}$. These distributions can be seen in Figure \ref{fig: age-vel}(left column) We report the median and MAD  of these distributions below. 

We find that the velocity dispersion of host stars of CCJ and CEJ is similar, $47.5\pm 2.3$ km/s and $47.9\pm 2.6$ km/s respectively. HJ hosts have smaller velocity dispersion ($40.6\pm 1.9$ km/s). On the other hand, WJ hosts have a velocity dispersion in between ($45.1\pm 1.9$ km/s). The smaller velocity dispersion indicates that HJ hosts are younger compared to CJ and WJ (See Figure \ref{fig: age-vel}). 

This velocity dispersion can be used as an age estimator. Following \citep{Narang22} we use \citet{AumerBinney2009} to find the average age ($\tau$) of an ensemble of stars from the total velocity dispersion as: 
\begin{equation}\label{eq: age}
    \tau = (10 \text{Gyr}+\tau_1)\left(\dfrac{\sigma_{tot}}{\nu_{10}}\right)^{(1/\beta)}-\tau_1
\end{equation}
From \citet{AumerBinney2009} : $\nu_{10}=57.157 $ $\text{km s}^{-1}$, $\beta=0.385$ and $\tau_1=0.261$ Gyr.
Using Equation \ref{eq: age}, we convert the distribution of $\sigma_{tot}$ to the corresponding age distributions (See Figure \ref{fig: age-vel}). We report the median and MAD of those age distributions below as MED $\pm$ MAD. 

CCJ and CEJ hosts have a median age of $6.07\pm0.79$ Gyr, and $6.25\pm 0.92$ Gyr respectively. Hot Jupiters are slightly younger, $3.97\pm0.51$ Gyr, and warm Jupiters have an average age of $5.28\pm0.6$ Gyr. Therefore, despite having a similar [Fe/H] on average, CEJ and HJ hosts have different average ages.

\section{Discussion} \label{sec:dis}
We chose a well-curated sample of main-sequence stars hosting Jupiter-like planets in the solar neighborhood with reliable measurements of metallicity([Fe/H]) and kinematics (proper motion, parallax, and radial velocity) from \textit{Gaia} DR3. Based on the orbital distances from their host star, and orbital eccentricity, we subdivided these Jupiters into 4 groups (HJ, WJ, CEJ, and CCJ). We have compared the host star age and metallicities of these groups.   

To summarise the results of the last sections, we find that, on average, HJ hosts are metal-rich ([Fe/H]=$0.18\pm0.13$) and young (Average age $\sim3.97\pm 0.51$ Gyr). On the other hand, CCJ hosts are relatively metal-poor, around solar metallicity ([Fe/H]=$0.03\pm 0.18$), and relatively older (average age $\sim 6.07\pm 0.79$ ). However, CEJ hosts, despite being metal-rich on average ([Fe/H]=$0.15\pm 0.12$) are relatively older (average age $\sim 6.25\pm 0.92$ Gyr).

The outcomes of our statistical tests on the [Fe/H] distributions of host stars unveil that HJ hosts and CCJ hosts do not belong to the same population. Similarly [Fe/H] distributions of CEJ and CCJ hosts most likely do not come from the same distribution. On the other hand, CEJ and HJ hosts likely come from the same population. However, we see an age difference between CEJ and HJ hosts, despite their metallicity being similar. 

\subsection{Implications of the results}\label{sec: impli}
As we have discussed in the introduction (Section \ref{sec:intro}), the correlation between orbital properties and host-star metallicities of giant planets has deep implications for the understanding of the formation and evolution pathways of giant planets. Here we discuss how our findings favor or disfavor each scenario. 
%How did the Jupiters form? Did they form in-situ? Or have they migrated from a distant origin to their current position? In the case of migration, which one is the dominant mechanism? Is it disk migration or high eccentricity tidal migration? All these are open questions, without a clear understanding. We discuss how our results favor or disfavor each scenario.
\subsubsection{In-situ formation}\label{sec: insitu}

The plausibility of forming hot and warm Jupiters at their present location is one of the major open questions \citep[e.g.,][]{Dawson18, Batygin2016, LeeChiang2017, Poon2021}{}{}. In order to form a Jupiter, a massive core needs to be formed before disk gas dissipates. However, the inner disk itself have a small amount of dust mass, and to form Jupiter the solid mass of the inner disk should be enhanced by two orders of magnitude \citep[e.g.,][]{Dawson18}{}{}. This enhancement should also reflect in metallicity of the host star. As a result, if in-situ formation is at play, hot Jupiter hosts should be very metal-rich on average. However, if we go further from the host star, massive core can grow within disk dispersal time even in disks with gradually smaller solid mass \citep[e.g.,][]{Mordasini12, Piso2015}. Therefore, Jupiters located in further orbits should have host stars with gradually decreasing average metallicity, on average. 
If the in-situ formation was the dominant mechanism for the formation of Jupiters, we should have seen a gradual decrease of the median metallicity of Jupiter hosts with orbital period \citep{Maldonado18}. Although we see hot Jupiter hosts to be metal-rich, we do not notice any significant difference in metallicity between warm and cold Jupiter hosts. In addition, the in-situ formation scenario can not account for the similarities in metallicity between CEJ and HJ hosts.

\subsubsection{Disk migration}\label{sec: diskmig}
In gas-disk migration, a massive planet perturbs the nearby gas and sends it onto horseshoe orbits via corotation torques, deflects distant gas by Lindblad torques, and exchanges angular momentum in the process \citep{Goldreich1980, Lin1986, Baruteau2014}. As a result, a net inward torque acts on the planet and the planet starts to migrate inwards. However, this mechanism is independent of the metallicity of the system \citep[e.g.,][]{Dawson18}. Therefore, if gas-disk migration were the dominant mechanism to sculpt all the close-in hot and warm Jupiters, we would not have seen a metallicity and orbital period correlation. On the other hand, gas-disk migration can not excite the eccentricity of planetary orbit to a high value\citep[e.g.,][]{Duffell2015}. Therefore warm Jupiters in eccentric orbits (See Figure \ref{fig:all3ecc}) are hard to explain by this mechanism. 

\subsubsection{High-eccentricity tidal migration}\label{sec: hemtides}
The most popular theory of forming HJs is by tidal migration of CEJs \citep[e.g.,][]{Socrates2012, Petrovich2015, Hamers2017, Dawson18, Teyssandier2019}. Above a threshold eccentricity, the CEJs can experience rapid tidal decay of their orbit. The first step is to excite the giant planets to high eccentricity. This can be attained if a giant planet interacts with a perturber. The perturber can be another giant planet or an external fly-by. If the excitation is due to an external fly-by, we do not expect any dependence on host-star metallicity\citep[e.g.,][]{Shara2016}. However, in metal-rich disks, giant planet formation is very efficient, due to the larger amount of solid mass present \citep[e.g.,][]{IdaLin2005, Mordasini12, Bitsch2015}. If multiple giants are formed in sufficiently close orbits, it may lead to planet-planet scattering, and one of them can be excited to very high eccentricity \citep[e.g.,][]{Rasio1996, Ford2008, Chatterjee2008}. Therefore, the metallicity enhancement of host stars of CEJ might be the indicator of planet-planet scattering happening in such initially metal-rich disks \citep[e.g.,][]{Dawson18, Buchhave18}.

If the eccentricity is very high, and the tidal migration timescale is very short, the CEJs tidally circularize in a shorter orbit, and end up as HJs \citep[e.g.,][]{Petrovich2015}. But if the eccentricity is below a threshold value, and the tidal migration timescale is comparable to the stellar lifetime. In that case, the CEJs remain in their eccentric orbit, with a large semi-major axis \citep[e.g.,][]{Petrovich2015}. 

This theory predicts that CEJs are progenitors of close-in HJs \citep[e.g.,][]{Socrates2012, Petrovich2015}. Therefore we would expect similar host-star properties between HJs and CEJs, on average. We do find that the [Fe/H] distribution of HJ and CEJ hosts are similar, and the null hypothesis of them being drawn from the same parent population could not be rejected. Both of the host star populations are also metal-rich, as we expect for planet-planet scattering to occur \citep[e.g.,][]{Chatterjee2008, Bitsch2015}.  

\begin{figure}
    \centering
    \includegraphics[width=0.4\textwidth]{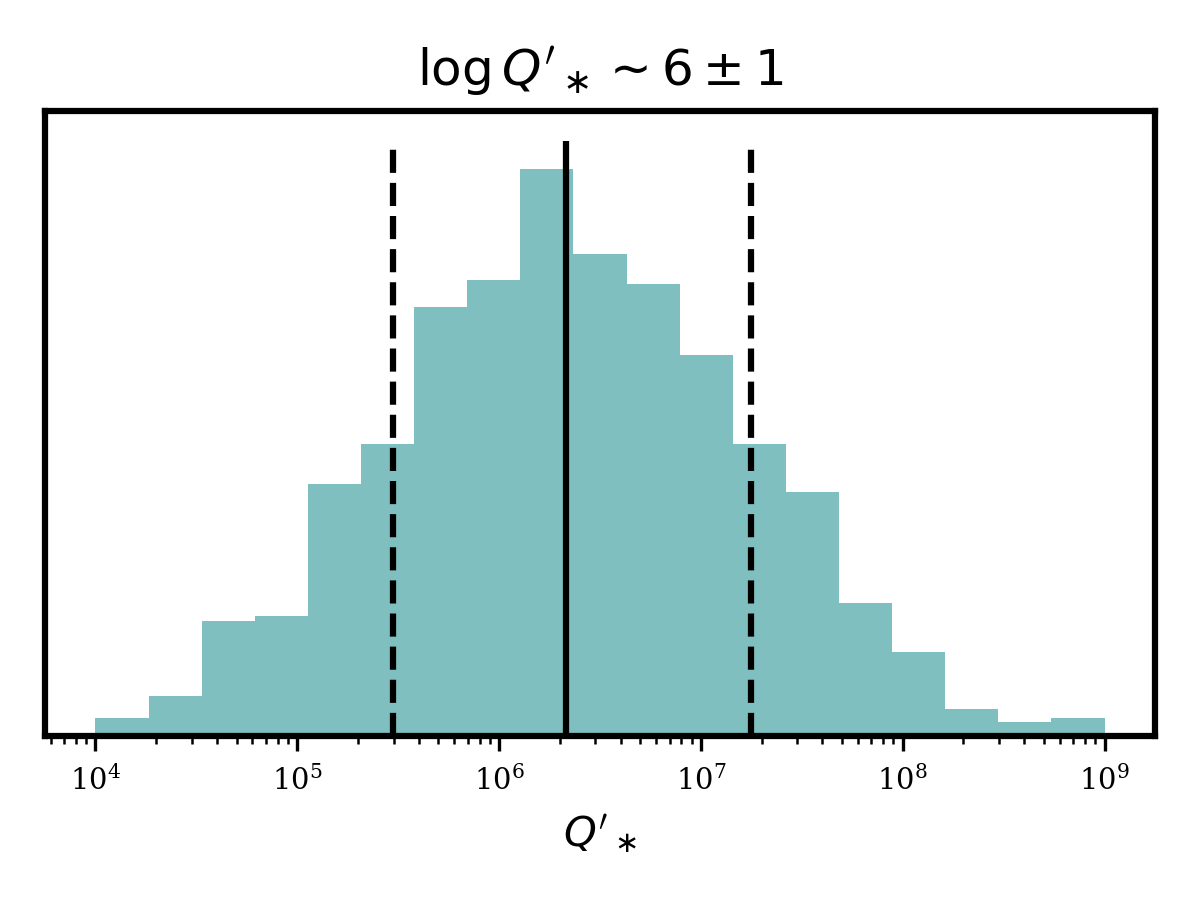}
    \caption{Possible distribution of $\log{Q'_\ast}$. If we assume inspiral timescale is $\sim 6$ Gyr, and the HJs that have been destroyed had similar planetary and stellar properties. The median, 16th and 84th percentile of $\log Q'_\ast\approx 6\pm 1$} 
    \label{fig:Q}
\end{figure}

On the other hand, we find that the average ages, derived from the velocity dispersion of HJ and CEJ hosts are different. HJ hosts are relatively younger than the CEJ hosts. But if CEJs indeed are the progenitors of HJ, the ages of their host stars should be similar. 

One possible explanation of our findings might be the destruction of older HJs  \citep[e.g.,][]{Hamer2019,Miyazaki2023}. If a significant fraction of older HJs are engulfed by the star, the remaining HJ hosts would appear younger on average. Even after orbital eccentricity is sufficiently damped, the HJs can fall into the host star by tidal interaction. The semi-major axis of a circularized Jupiter shrinks because of the tides raised by the planet on the star \citep[e.g.,][]{Goldreich1966, Jackson2008, Barker2020}. The infall timescale ($t_{\text{in}}$), for a HJ with orbital period $P$, semi-major axis $a$ and mass $M_p$ around a star of mass $M_\ast$ is given by \citep[e.g.,][]{Lai2012,Barker2009,Hamer2019}:

\begin{equation}\label{eq: tin}
    t_{\text{in}}=\dfrac{2}{13}\left|\dfrac{a}{\dot{a}}\right|=\dfrac{2}{13}\dfrac{2Q'_\ast}{9}\dfrac{M_\ast}{M_p}\dfrac{P}{2\pi}\left(\dfrac{a}{R_\ast}\right)^5
\end{equation}

Where $Q'_\ast=3Q_\ast/2k$ is the modified stellar tidal quality factor. $Q_\ast$ is the ratio of the maximum energy stored in tides to the energy dissipated in one orbital cycle, and $k$ denotes the tidal Love number. 
%We note that, the median age difference between CEJ hosts and HJ hosts is {}. The median period and mass of the HJs in our sample is {} and {} respectively. If we equate the median age difference to the inspiral timescale, we get an estimate of $Q'_\ast\sim {}$.  

The value of $Q'_\ast$ depends on the internal structure of the star, tidal forcing frequency and amplitude in general \citep[e.g.,][]{Barker2009,Barker2020,Miyazaki2023}{}{}. There have been attempts to constrain the value of $Q'_\ast$ from theory \citep[e.g.,][]{Barker2009,Penev2018} and observed datasets \citep[e.g.,][]{Bonomo2017,Labadie2019,Hamer2019,Miyazaki2023}. Assuming CEJs are the true progenitors of HJs, we try to estimate $Q'_\ast$ of Jupiter hosts in the following way. 

We note from Figure \ref{fig: age-vel} that median age of CEJ hosts is $\sim 6$ Gyr, and the upper limit of the age of the HJ hosts is $\sim$ $6$ Gyr. Therefore, the probability of finding a HJ host older than 6 Gyr is very small. If CEJs are progenitors of HJs, and in a timescale of $\sim 6$ Gyr most of the HJs are engulfed by the star, this scenario may occur. 

If we equate tidal inspiral timescale ($t_{\text{in}}$) to $6$ Gyr, we will get an estimate of $Q'_\ast$, depending on stellar and planetary parameters. However, we note that, for a CEJ to become an HJ and then to be engulfed by the star, two processes are involved. (1) High-eccentricity tidal migration of CEJs to a shorter orbit (2) Tidal inspiral of HJ onto the star. Figure \ref{fig: age-vel} indicates that sum of the two timescales should be $\sim 6$ Gyr. We are assuming that $t_{\text{in}}$ is much larger compared to high-eccentricity tidal migration timescale. In other words, we are placing an upper limit on $Q'_\ast$.

If we assume the HJs that have been engulfed by the star had similar planetary and host star properties to the observed HJs, estimation of $Q'_\ast$ is straight forward. We draw 1000 random points from the planetary mass, orbital period, stellar radius and stellar mass and plug into Equation (\ref{eq: tin}) to get a distribution of $Q'_\ast$. The distribution of $Q'_\ast$ is shown in Figure (\ref{fig:Q}). We find the median, 16th and 84th percentile
\begin{equation}
   \log{Q'_\ast}\approx 6\pm1
\end{equation}

This range of values of $Q'_\ast$ is consistent with those existing in literature \citep[e.g.,][]{Penev2018, Hamer2019, Miyazaki2023} and can lead to the destruction of a significant fraction of the older HJ population, leading to the average young age, and an average age difference with the CEJs. 

However, this value is only an order of magnitude estimate of $Q'_\ast$, and holds only in the premise of our assumptions. To explain the stability of short-period HJs, \citet{Lai2012} argues for a larger value of $Q'_\ast \sim 10^8-10^9$. 
%We also note that assuming a uniform value of $Q_\ast$ for all the systems might be unrealistic. But for a fraction of HJ to tidally infall within a timescale to create an observable difference between the average age of CEJ and HJ hosts, our analysis shows, that a smaller value of $Q_\ast$ is needed. 

\begin{figure}[htb!]
    %\begin{subfigure}{0.49\textwidth}
    \subfloat{
        \centering
        \includegraphics[width=0.45\textwidth]{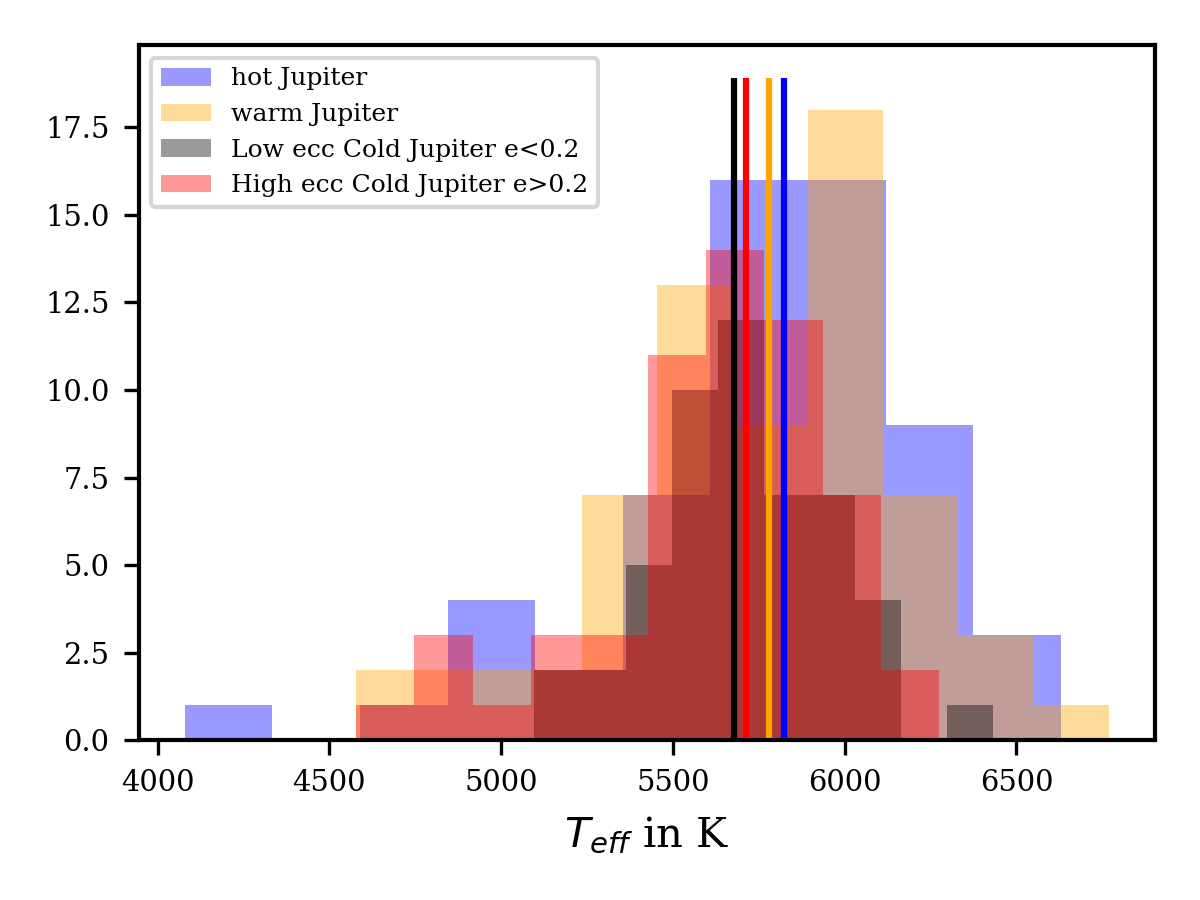}
        \label{fig:teff1}}
    \vfill
    %\end{subfigure}
    %\begin{subfigure}{0.49\textwidth}
    \subfloat{
        \centering
        \includegraphics[width=0.45\textwidth]{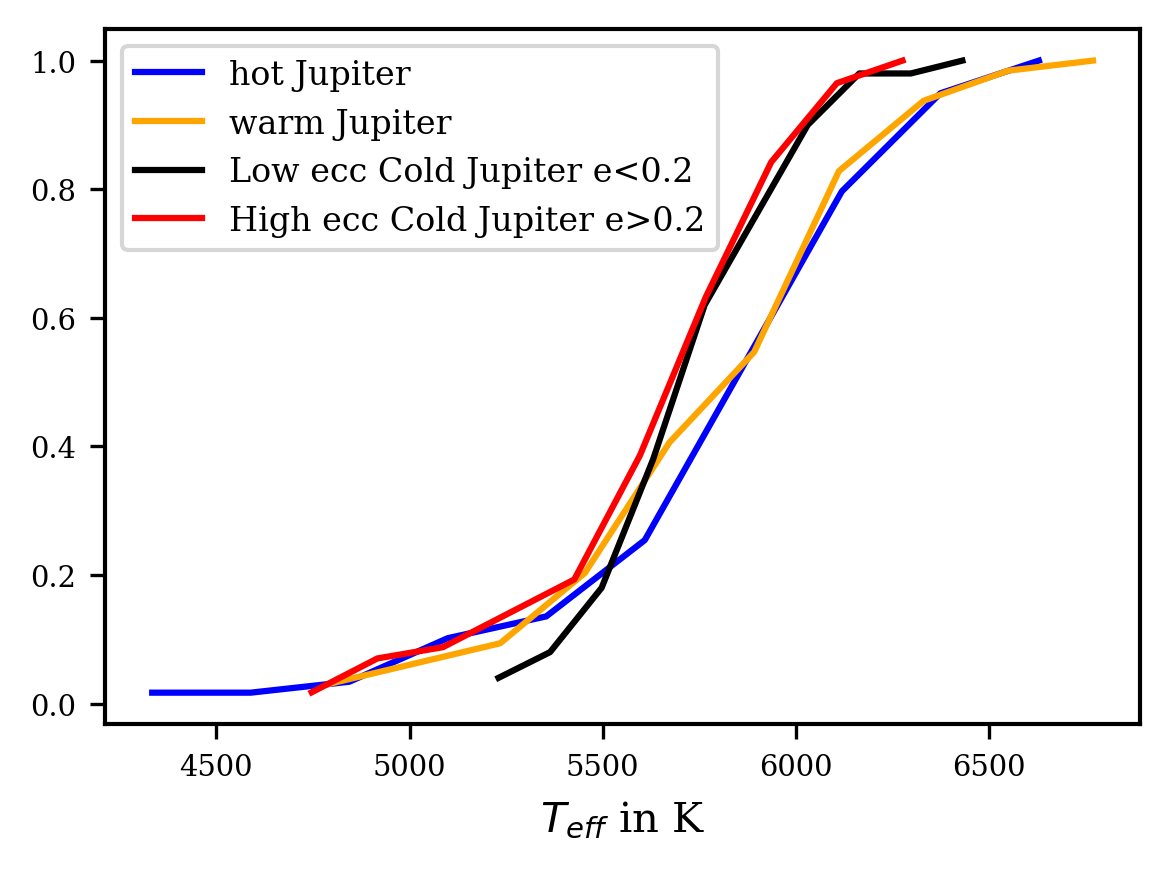}
        \label{fig:teff2}}
    %\end{subfigure}
    \caption{$\text{T}_\text{eff}$ distribution of host stars of HJ, WJ, CEJ, and CCJ. In terms of $\text{T}_\text{eff}$ we do not see a significant difference.}
    \label{fig:teff}
\end{figure}

 For consistency check, and to check for possible bias in the sample, we also compared the effective temperature ($\text{T}_\text{eff}$) of the host stars of HJ, WJ, CEJ and CCJ ( Figure \ref{fig:teff}). It is apparent from Figure \ref{fig:teff} that in terms of $\text{T}_\text{eff}$ we do not see a significant difference between the host stars of different groups. We find that HJ and WJ hosts are marginally hotter, with $\delta T_\text{eff}< 200$K, and CEJ and CCJ hosts have similar $\text{T}_\text{eff}$ on average (Figure \ref{fig:teff2}). We note that the observed metallicity-age distribution can not be due to this marginal difference of $\text{T}_\text{eff}$. 

\section{Summary}\label{sec: sum}
\begin{enumerate}
    \item We started with a well-curated sample of 702 planet hosts in main-sequence within 250 pc, with metallicity ([Fe/H]) and kinematics homogeneously measured and reported in \textit{Gaia} DR3. We have used only the best quality data products from \textit{Gaia} DR3 following \citet{Recio-Blanco22}. We made consistency checks of \textit{Gaia} DR3 metallicities with \textit{GALAH} and \textit{LAMOST}. For our analysis we used only Jupiter hosts, defining Jupiter as a planet with mass($\text{M}_\text{J}$) between ($100M_\oplus<\text{M}_\text{J}<1200M_\oplus$).
    \item We subdivided the Jupiters into three groups based on their orbital period, HJ, WJ, and CJ; and compared host star properties, namely metallicity ([Fe/H]) and age, of these three groups. We find that HJ hosts are the more metal-rich ([Fe/H] =$0.18\pm 0.13$) while WJ and CJ hosts have similar metallicity on average ([Fe/H] =$0.08\pm0.18$ ). Most of the HJs are in circular orbit, but we also find, a significant fraction of WJ and CJ are in eccentric orbits. We find no difference in host star metallicity between low and high eccentric WJs. However, for CJs, the eccentric population (CEJ) has high host star metallicity ([Fe/H] =$0.15\pm0.12$) on average compared to the low eccentric population (CCJ). CCJ hosts have [Fe/H] close to solar value ([Fe/H] =$0.03\pm0.18$). This finding, from a larger and homogeneous dataset, agrees with previous work in literature by \citet{Buchhave18}.
    %We found that the host star metallicity ([Fe/H]) and orbital eccentricity of CJ has a bimodality. The eccentric population, CEJ, are preferentially found around more metal-rich host star on average. While the cold Jupiters on circular orbits, have host stars with solar metallicity on average. 
    \item To test the statistical significance of the observed result, we have performed Monte Carlo Analysis and several non-parametric statistical tests, as described in Section \ref{sec:statest}. Our findings suggest that the observed difference in host star metallicity between HJ and CCJ, and between CEJ and CCJ, can not be a consequence of random sampling from an underlying parent population. Based on the [Fe/H] distributions of host stars, we can conclude, HJ and CCJ, as well as CEJ and CCJ, do not come from the same parent population. However, HJ and CEJ likely come from the same underlying population, and the null hypothesis can not be ruled out. In other words, CEJs might be progenitors of HJ in very metal-rich systems. Whereas, CCJs are born separately, in relatively metal-poor environments.  
    
    \item  We find host stars of HJs and CEJs have similar [Fe/H] on average, but their average age is different. The similarity in [Fe/H] distribution supports the theory of high eccentricity migration of CEJ as the progenitors of HJs, triggered by planetary perturbation \citep[e.g.,][]{Buchhave18, Bitsch2015}. However, the difference in average age indicates that older HJs might be getting destroyed, hence the HJs appear younger on average. If we assume CEJ are indeed progenitors of HJs and a fraction of HJs are getting destroyed because of the tidal interaction with the star, we find the average value stellar modified tidal quality factor ($Q'_\ast$) to be: ${Q'_\ast}\approx10^{6\pm 1}$. This range of values agrees with other works in literature \citep[e.g.,][]{Hamer2019, Penev2018, Miyazaki2023}.
\end{enumerate}

\section{Acknowledgement} \label{sec: ack}

%\section{Appendix}\label{sec:app}
We thank Prof. Thomas Henning for the insightful suggestions. The authors thank the anonymous referee for valuable feedback and useful insights. This research has made use of the NASA Exoplanet Archive \citep{exoarch}, which is operated by the California Institute of Technology, under contract with the National Aeronautics and Space Administration under the Exoplanet Exploration Program. This work has made use of data from the European Space Agency (ESA) mission
{\it Gaia} (\url{https://www.cosmos.esa.int/gaia}), processed by the {\it Gaia}
Data Processing and Analysis Consortium (DPAC,
\url{https://www.cosmos.esa.int/web/gaia/dpac/consortium}). Funding for the DPAC
has been provided by national institutions, in particular, the institutions
participating in the {\it Gaia} Multilateral Agreement. We have made use of several packages from \texttt{Scipy} \citep{Scipy2020} for statistical tests. We have used the \texttt{emcee} \citep{emcee}  We sincerely thank the Infosys Foundation for funding the travel to PPVII in Kyoto to present this work as a poster. We acknowledge support of the Department of Atomic Energy, Government of India, under Project Identification No. RTI 4002. 
\appendix
\section{Non-parametric statistical tests}\label{sec: appen}
If the true nature of the original distribution is unknown, the best way to compare two samples is to make use of non-parametric statistical tests. In this work, we have used several tests, here we describe them briefly. For details please see \citet{nonpar}. 

\textbf{1. Mann-Whitney (MW) U test:}
This test compares between central tendencies of two distributions. In this test, two samples are merged and the combined sample is sorted in ascending order and ranked \citep[e.g.][]{MannWhitney47, nonpar}. The strategy is to see if the ranks are randomly mixed or ranks of two groups are clustered at opposite ends. The test computes the individual sums of the ranks of the two samples. The smaller of the two sums is called the U statistic. For large samples (at least $>8$ for each group) the distribution of U statistic can be approximated with a normal distribution. The test then evaluates whether the observed U statistic significantly deviates from what would be expected under the null hypothesis of no difference between the distributions. 

\textbf{2. Kruskal-Wallis (KW) test:}
Kruskal-Wallis (KW) H test \citep[e.g.,][]{kruskalWallis52,nonpar} tests the null hypothesis that the population medians of all the data groups are equal. It is a non-parametric extension of the ANOVA (Analysis of Variance) test. This test is used to compare two or more samples. First, all the samples are combined and ranked in the ascending order. For comparison between k groups, The KW H statistic is defined by:
\begin{equation*}
    H=\dfrac{12}{N(N+1)}\sum^k_{i=1}\dfrac{R_i^2}{n_i}-3(N+1)
\end{equation*}
Where $R_i$ is the sum of the ranks from a particular group, and $n_i$ is the number of values from the corresponding rank sum. 
If there are ties between the ranks, the statistic is divided by the following correction ($C_H$), and a new statistic is calculated. 
\begin{equation*}
    C_H=1-\dfrac{\sum (T^3-T)}{N^3-N}
\end{equation*}
If the sample size of the individual groups is large, KW statistic asymptotically approaches the chi-squared distribution. In our case, all groups have large samples ($>50$) for this assumption to hold. The p-value of the test is then obtained by computing the survival function from the chi-square distribution at H. 

\textbf{3. Kolmogorov-Smirnov (KS) test:}
In this test \citep[e.g.,][]{Smironov1948, Hodges1958, nonpar}, two empirical distribution functions (e.d.f)s are constructed from two datasets, and the maximum vertical distance between two functions is computed. If we have two samples of size m and n from cumulative distribution function(c.d.f)s F and G and wish to test the null hypothesis that $F(x)=G(x)$ for all x, the two sample KS statistic is defined by:
\begin{equation*}
    D_{KS}=max |\hat{F}_m(x)-\hat{G}_n (x)|
\end{equation*}
where $\hat{F}_m, \hat{G}_n$ are two e.d.f's. 
The null hypothesis is rejected at confidence level $\alpha$ if 
\begin{equation*}
    D_{KS} > c(\alpha)\sqrt{\dfrac{n+m}{n. m}}
\end{equation*}
Where $c(\alpha)=\sqrt{-0.5\ln{(\alpha/2})}$ in general. 
\bibliography{sample631}{}
\bibliographystyle{aasjournal}



\end{document}